\documentstyle[aps,twocolumn,psfig,epsf]{revtex}
\begin{document}
\draft
\title{\large \bf RELATIVISTIC INSTANT--FORM APPROACH TO
THE STRUCTURE OF
TWO-BODY COMPOSITE SYSTEMS. II. NONZERO SPIN}

\author{A.~F.~Krutov\thanks{Electronic address:
krutov@ssu.samara.ru}}

\address {\it Samara  State University, RU-443011, Samara,
Russia}

\author{V.~E.~Troitsky\thanks{Electronic address:
troitsky@theory.sinp.msu.ru}}
\address {\it D.~V.~Skobelltsyn Institute of Nuclear Physics ,
Moscow State University, RU-119899, Moscow, Russia}

\date{September, 2002}
\maketitle
\begin{abstract}
The relativistic approach to electroweak properties
of two-particle composite systems developed in  Ref.\cite{KrT02}
is generalized here to the case of nonzero spin.
This approach is based on the use of
the instant form of relativistic Hamiltonian dynamics.
The generalization makes use of a special mathematical technique
for the parametrization of matrix elements of electroweak
current operators in terms of form factors.  In
this technique the parametrization is a realization of the
Wigner--Eckart theorem on the Poincar\'e group and form factors
are reduced matrix elements. As in the case of
zero spin the electroweak current matrix element satisfies the
relativistic covariance conditions and in the case of
electromagnetic current it also automatically satisfies the
conservation law. Physical approximations such as, for example,
the relativistic impulse approximation, are formulated in terms
of reduced matrix elements. The electromagnetic structure of
$\rho$ meson is calculated as an example of realization of the
technique proposed.  \end{abstract}

PACS number(s): 13.40.--f, 11.30.Cp
\narrowtext

\section{Introduction}

A new relativistic approach to electroweak properties of
composite systems has been proposed in our recent paper
\cite{KrT02}. The approach is based on the use of the instant form (IF)
of relativistic Hamiltonian dynamics (RHD).
The detailed description of RHD can be found in the review
\cite{KeP91}.
Some other
references as well as some basic equations of RHD
approach are given in  Ref.\cite{KrT02}.

In the paper  \cite{KrT02} our approach was used to perform a
realistic calculation of electroweak properties of pion considered as
composite quark--antiquark system. The electromagnetic form factor and the
lepton decay constant were calculated for pion using different
model wave functions for the relative motion of quarks in pion.

Now our aim is to generalize the approach to more complicated
systems, namely, to composite systems of two particles of
spin 1/2 with nonzero values of total angular momentum,
total orbital momentum and total spin.  The main problem is
a construction of electromagnetic current operator satisfying
standard conditions (Lorentz covariance, conservation law
etc., see, e.g., Ref.\cite{KrT02}).

The basic point of our approach
\cite{KrT02} to the construction of the electromagnetic current
operator is the general method of relativistic invariant parameterization of
local operator matrix elements proposed as long ago as in
1963 by Cheshkov and Shirokov~\cite{ChS63}.
This canonical parametrization of local
operators matrix elements was generalized to the case of composite systems
of free particles in Refs.~\cite{TrS69,KoT72}.

In fact, this parametrization is a
realization of the Wigner--Eckart theorem for the Poincar\'e
group and so it enables one for given matrix element of arbitrary
tensor dimension to separate the reduced
matrix elements (form factors) that are invariant
under the Poincar\'e group. The matrix element of a
given operator is represented as a sum of terms, each one of
them being a covariant part multiplied by an invariant part.
In such a representation a covariant part describes
transformation (geometrical) properties of the matrix element,
while all the dynamical information on the transition is
contained in the invariant part -- reduced matrix elements.
In the case of composite systems these form
factors appearing through the canonical parameterization are to
be considered in the sense of distributions, that is they are
generalized instead of classical functions. As was demonstrated
in Ref.\cite{KrT02} this fact takes place even in nonrelativistic
case. It is in terms of form factors that the electroweak
properties of composite systems are described in the frame of
the approach \cite{KrT02}.

In our approach some rather general problems arising in the
description of composite quark models have been solved. For example,
the description of electromagnetic properties of composite
systems in terms of form factors in Ref.\cite{KrT02}, in fact,
solves the problem of construction of the electromagnetic
current satisfying the conditions of translation invariance,
Lorentz covariance, conservation law, cluster separability and
equality of the composite system charge to the sum of constituents charges
(charge nonrenormalizability).

Let us note that the importance of the problem of the
construction of the electromagnetic current is
actual not only for RHD but for all relativistic approaches to
composite systems, including the field theoretical approaches
\cite{Lev95,GrR87,ChC88,Kar92,VaD95,LeP98,MeS98,Kli98}.

We will construct the electromagnetic current operator in the
frame of IF of RHD. The same problem was considered in
Refs.~\cite{Lev95,Kli98} in the point form of RHD and in
Refs.~\cite{LeP98,MeS98} in the case of light--front dynamics.

Our approach is a generalization of the method \cite{Kli98} to
the case of the instant form dynamics. However, the scenario of
the generalization of the Wigner--Eckart theorem is quite
different.

Physical approximations that we use in our approach are
formulated in terms of reduced matrix elements, for example, the
well known relativistic impulse approximation.
It means that the electromagnetic current of a composite system
is a sum of one--particle currents of the constituents. It is
worth emphasizing that in our method this
approximation does not violate the standard conditions
for the current listed before. To--day a construction of
relativistic impulse approximation without breaking of
relativistic covariance and current conservation law is a common
trend of different approaches
\cite{Lev95,VaD95,LeP98,Kli98}.
Let us note that in the present paper it is for the first time
that such a construction has been realized for the case of nonzero
spin in the frame of IF RHD. This is a variant of the
relativistic impulse approximation (IA) formulated in terms of
reduced matrix elements (see Ref.~\cite{KrT02}) -- the modified
impulse approximation (MIA).

The canonical parameterization of electroweak current matrix
element, i.e. the extraction of the reduced
matrix elements in the case of zero total spin and zero total
angular momentum was performed in Ref.~\cite{KrT02} and is
rather simple.  The case of composite systems with nonzero
values of total spin and angular momentum requires the
development of a general method for canonical parameterization
of local operator matrix elements. Here we develop an adequate
mathematics using as a base the paper \cite{ChS63}.

In the present paper we propose a general formalism for the
operators diagonal in the total angular momentum. The case of
non-diagonal operators describing, for example,
the radiative transitions between vector and scalar composite particles
will be considered elsewhere. We demonstrate the application of
the formalism in the case of a system with total spin one and
total angular momentum one  and with zero orbital momentum. In
this connection the $\rho$ -- meson electromagnetic structure is
calculated as an example. Using different model wave functions
of the quark relative motion we calculate the electromagnetic
form factors and the static properties of the $\rho$ meson
supposing quarks to be in the $S$ state of relative motion. It
is interesting to mention that relativistic effects occur to
produce a nonzero quadrupole momentum and quadrupole form
factor. It is well known that in the nonrelativistic case the
non--zero quadrupole form factor is caused by the presence of
the $D$- wave and is zero otherwise.

It is worth noticing that our approach guarantees the uniqueness
of the solution for the
$\rho$ -- meson electromagnetic form factors in contrast with the
calculations based on the standard light--front dynamics
in Refs.~\cite{Kei94,CaG95pl} where it is shown that the
form factors obtained from matrix elements with different
total angular momentum projections differ from one another themselves.
An analogous ambiguity takes place in the light-front dynamics
calculations of the deuteron electromagnetic structure
\cite{GrK84}.
The next step of generalization of our approach concerns
two--particle systems with nonzero orbital momentum giving a
possibility to describe this simplest nuclear composite
system~\cite{KrTxx}.

The paper is organized as follows.
In Sec.II the canonical parametrization of local operator matrix
elements between one--particle states of arbitrary nonzero spin
is described. This parametrization presents the extraction of
the reduced matrix element on the Poincar\'e group. The
electromagnetic current matrix element is derived for the spin
1/2 particle. For this case the relations between the form
factors in the canonical parametrization form and commonly used
Sachs form factors are given explicitly.

In Sec.III we show how to construct the electromagnetic current
operator for composite system of two free particles of spin 1/2.
The current matrix element is obtained in the basis where the
center--of--mass motion is separated. The electromagnetic
properties of the system are defined by reduced matrix
elements, or the so called free two--particle form factors,
these form factors being generalized functions. This means
that, for example, the static properties of the system are given
by the weak limits as $Q^2\to$ 0
($Q^2 = -q^2\;,\;q$ is the momentum transfer).
A special attention is paid to the case of total spin
one and zero total orbital momentum. In
this case the electromagnetic properties are defined by four
free two--particle form factors -- charge, quadrupole,
magnetic and the magnetic quadrupole form factor of the second kind
(see the review in Ref.~\cite{DuT90} and the references therein).

In Sec.IV the electromagnetic current matrix element is
constructed for the case of two interacting particles. Each step
of the calculation in the developed method
remains Lorentz covariant and
conserves the current. The composite system form factors are
derived as some integral representations in terms of the wave
functions obtained in the frame of RHD.

In Sec.V the developed formalism is used in the case of the
system with total spin one and total angular momentum one. In
terms of reduced matrix elements the so called modified
impulse approximation (MIA)
~\cite{KrT02} is formulated. In MIA
$\rho$ -- meson form factors are obtained explicitly.

In Sec.VI the results of calculations of static properties
and electromagnetic form factors of
$\rho$ meson are discussed.

In Sec.VII the conclusion is given.

\section{Parametrization of one--particle operator
matrix elements}

A parametrization of the current operator matrix element is
a representation of the matrix element in terms of form factors.
Let us remark that it is just form factors -- Lorentz invariant
functions -- are extracted from scattering data.
In simple cases, such as particles with spin 0 or 1/2
the parametrization can be obtained through semiempirical
approaches. However, in more complicated cases, for example, in
the case of nuclei with arbitrary spins one needs a general
mathematical method.

Let us describe now a general method for canonical
parameterization of local operator matrix elements ( see
Ref.~\cite{ChS63}, too). As we have mentioned before, here we are
dealing with the matrix elements diagonal in total angular
momentum only. Non-diagonal case will be considered elsewhere.

From the group theory point of view a parameterization is, in
fact, a realization of the well known Wigner--Eckart theorem on
the Poincar\'e group. The parameterization extracts the reduced
matrix elements that are invariant under the
Poincar\'e group. These reduced matrix elements
are just form factors that are measured in
experiments.

The main idea of the parametrization can be formulated as
follows. Using the variables entering the state vectors
which define the matrix elements one has to construct two
types of objects.

1. A set of linearly independent matrices which are
Lorentz scalars (scalars or pseudoscalars). This set describes
transition matrix elements non-diagonal in spin projections
in the initial and finite states, as well as the properties
defined by the discrete space--time transformations.

2. A set of linearly independent objects with the same
tensor dimension as the operator under consideration
(for example, four--vector, or four--tensor of some rank). This
set describes the matrix element behaviors under the action of Lorentz group
transformations. In the case of Lorentz scalar
operator the second set coincides with the first one.

The operator matrix element is written as a sum of all possible
objects of the first type multiplied by all the possible objects
of the second type. The coefficients in this representation as a
sum are just the reduced matrix elements -- form factors. The
obtained representation is then modified with the use of
additional conditions for the operator, such as the conservation
laws, for example. In order to satisfy these additional conditions in
some cases some of the coefficients occur to be zero.

To demonstrate this let us consider the parameterization of the
matrix elements taken between the states of a free particle of
mass $M$ in different simple cases. Let us normalize the state
vectors as follows
~\cite{KrT02}:
\begin{equation}
\langle\,\!\vec p\,,m\,|\,\vec p\,'\,,m'\,\!\rangle =
2p_0\,\delta (\vec p - \vec p\,')\,\delta _{mm'}\;,
\label{normg}
\end{equation}
Here $\vec p\;,\;\vec p\,'$ are three--momenta of particle,
$p_0 = \sqrt{M^2 + \vec p\,^2}$, $m\;,\;m'$ are spin
projections.

Let us consider first the parameterization of the matrix
element of scalar operator
$A(x)$ taken between the states of free particle with zero
spin. Because of translation invariance it is sufficient to
consider $A(0)$. Using the variables of state vectors in the
initial and final states one can construct two linearly
independent scalars:
$p^2 = {p'}\,^2 = M^2\;,\;(p - p')^2 = q^2 = - Q^2$.
Only the second one -- the momentum transfer
square -- is nontrivial. Ignoring the trivial dependence on $M$ we can write:
\begin{equation}
\langle\,\vec p,\,M|A(0)|\,\vec p\,',\,M\,\rangle = f(Q^2)\;.
\label{A(0)}
\end{equation}

Let us consider now the matrix element of a scalar operator
between the states of a particle with spin $j$:
\begin{equation}
\langle\,\vec p,\,M,\,j,\,m\,|A(0)|\,\vec p\,',\,M,\,j,\,m'\,\rangle
\label{A(0)j}
\end{equation}

The tensor dimension of the operator is not changed,
nevertheless, the matrix element is now a matrix in spin
projections in the initial and finite states.

Let us construct the set of Lorentz scalars -- linearly
independent matrices in spin projections to be used for
the construction of the representation of Eq.~(\ref{A(0)j}).  Let
us use the covariant spin operator $\Gamma^\mu(p)$ \cite{Shi51}.
In the rest frame this operator coincides with the particle spin
operator:  
\begin{equation} 
\Gamma^0(0) = 0\;,\quad
\vec\Gamma(0) = M\,\vec j\;,\quad [\,j_i\,,j_k\,] =
i\,\varepsilon_{ikl}\,j_l\;.  
\label{Lspin} 
\end{equation} 
The covariant spin operator can be defined with the use of the
Pauli-Lubanski vector $w^\mu$ \cite{Nov72}. In terms of matrix
elements we have:  
$$ 
\Gamma^\mu_{mm'}(p) = \langle\,\vec p\,,m|w^\mu|\vec p\,,m'\rangle 
$$ 
\begin{equation} 
= \langle\,0\,,m|\hat U^{-1}(\Lambda_p)w^\mu \hat U(\Lambda_p)|0\,,m'\rangle
\label{Gw}
\end{equation}
$$
= \left(\Lambda_p\right)^\mu_\nu\langle\,0\,,m|w^\nu|0\,,m'\rangle
= \left(\Lambda_p\right)^\mu_\nu\,\Gamma^\nu_{mm'}(0)\;.
$$
Here $\Lambda_p$ is the boost for the transformation from 
the rest frame of the particle to the laboratory frame,
$\hat U(\Lambda_p)$ is the corresponding representation
operator. The matrix
$\Lambda_p$ can be written explicitly using the matrix for
Lorentz transformation of the vector $p'$ into the vector $p$:
\begin{equation}
\Lambda ^\mu _{\;\nu} = \delta^\mu_\nu + \frac{2}{M^2}p^\mu\,p'_\nu -
\frac{(p^\mu + {p'}\,^\mu)(p_\nu + p'_\nu)}{M^2 + p^\lambda\,p'_\lambda}\;.
\label{Lambda}
\end{equation}
In our case $\Lambda_p$ is given by Eq.~(\ref{Lambda}) with
$p'=(M\;,\;0\;,\;0\;,\;0)$.

Now equations (\ref{Lspin}) can be written as:
$$
\Gamma^0_{mm'}(0) = \langle\,0\,,m|w^0|0\,,m'\rangle = 0\;,
$$
\begin{equation}
\vec \Gamma_{mm'}(0) = \langle\,0\,,m|\vec w|0\,,m'\rangle =
M\,\vec j_{mm'}\;.
\label{Gj}
\end{equation}
Using the explicit form of $\Lambda_p$ we obtain:
$$
\Gamma_0(p) = (\vec p\vec j)\;,\quad
\vec \Gamma(p) =  M\,\vec j + \frac {\vec p(\vec p\vec j)}{p_0 + M}\;,
$$
\begin{equation}
\quad \Gamma^2 = -M^2\,j(j+1)\;.
\label{ Gamma mu}
\end{equation}

As is known \cite{Nov72}, spin transforms under the action of
Lorentz group following the so called little group which is
isomorphic to the rotation group, that is the corresponding
transformations are realized by the matrices of three
dimensional rotations.
The derivation of the explicit form of these matrices with
an arbitrary spin is described in Ref. \cite{Che66}.
For spin 1 and 1/2 the matrices have the form:
$$
D^{1/2} (p_1 ,p_2) = \cos\frac{\omega}{2} - 2\,i\,(\vec k\cdot\vec j)\,
\sin\frac{\omega}{2}\;,
$$
$$
D^{1}(p_1,p_2) = I - i( \vec k\cdot\vec j)\,\sin\omega +
( \vec k\cdot \vec j)^2\,(\cos\omega - 1)\;,
$$
$$
\omega = 2\,\arctan\,\frac {|[\, \vec p_1\,, \vec p_2\,]|}
{(p_{10} + M_1)(p_{20} +M_2) - ( \vec p_1 \vec p_2)}\;,
$$
\begin{equation}
\vec k = \frac {[\, \vec p_1\,, \vec p_2\,]}
{|[\, \vec p_1\,, \vec p_2\,]|}\;.
\label{Dj}
\end{equation}

So, the 4-spin operator is transformed under Lorentz
transformations $p^\mu = \Lambda^\mu_{\;\nu}\,p'\,^\nu$
in the following way:
\begin{equation}
\Gamma^\mu(p) = \Lambda ^\mu_{\;\nu}\,D^j(p,\,p')\,
\Gamma^\nu(p')\,D^j(p',\,p)\;.
\label{Lambda Gamma mu}
\end{equation}
Using Eq.~(\ref{Lambda Gamma mu})
one can show directly that the matrix elements of the operators
$D^j(p,\,p')\Gamma^\mu(p')$ and $\Gamma^\mu(p)D^j(p,\,p')$
transform as 4--pseudovectors, the matrix elements of the
operators $D^j(p,\,p')p_\mu\Gamma^\mu(p')$ and
$p'_\mu\Gamma^\mu(p) D^j(p,\,p')$ -- as 4-pseudoscalars.

We construct the set of linearly independent Lorentz--scalar
matrices using the vectors
$p^\mu,\;{p'}^\mu$ and the pseudovector
$D^j(p,\,p')\Gamma^\mu(p')$.
Note, that the pseudovector
$\Gamma^\mu (p)D^j(p,\,p')$ does not enter the
decomposition,
being linearly dependent. One can show this fact using the
relation (\ref{Lambda Gamma mu}) and the explicit form
Eq.~(\ref{Lambda}) of $\Lambda ^\mu _{\;\nu}$ transforming  $p'$
into $p$ .  It is easy to obtain:  
$$ 
\Gamma^\mu (p)D^j(p,\,p') = D^j(p,\,p')\left[\Gamma^\mu (p')\right.  
$$ 
\begin{equation}
\left. - \frac {p^\mu +{p'}^\mu}{M^2 + p_\mu {p'}^\mu}\,
\left[p_\nu \Gamma^\nu (p')\right]\right]\;.
\label{Gamma D = D Gamma}
\end{equation}
As ${p'}_\mu \Gamma^\mu (p') = 0$, the set in question of
linearly independent matrices in spin projections of
the initial and the final states giving the set of independent
Lorentz scalars is presented by $2j + 1$ quantities
\begin{equation}
D^j(p,\,p')\,(p_\mu \Gamma^\mu (p'))^n\;,\quad n = 0,1,\ldots ,2j\;.
\label{pseud}
\end{equation}
The number of linearly independent scalars in
Eq.~(\ref{pseud}) is limited by the fact that the product of more
than $2j$ elements
$\Gamma^\mu (p')$ reduces as it is linearly dependent.
If $n$ is even, then the obtained quantities are scalars, if $n$
is odd then they are pseudoscalars. The current matrix element (\ref{A(0)j})
is represented by the linear combination of these linearly
independent Lorenrz scalars. The coefficients in this
combination $f_n(Q^2)$ are just form factors.
These form factors are invariant under rotations and so they do
not depend on spin projections. So, they depend upon only one
scalar combination of variables -- the momentum--transfer squared.

For self--adjoint operator $A(0)$ a minor modification of the set
(\ref{pseud}) is necessary: in the scalar product in
Eq.~(\ref{pseud}) the factor $i$ appears.
So, now the current matrix element
(\ref{A(0)j}) can be written in the form:  
$$
\langle\,\vec p,\,M,\,j,\,m\,|A(0)|\,\vec p\,',\,M,\,j,\,m'\,\rangle 
$$ 
$$ 
= \sum _{n=0}^{2j}\sum_{m''=-j}^{j}\,\langle\,m|D^j(p,\,p')|m''\rangle 
$$
\begin{equation}
\times
\langle\,m''\,|\{ip_\mu\Gamma^\mu (p')\}^n|\,m'\,\rangle\,f_n(Q^2)
\label{DA(0)j= f_n}
\end{equation}
The obtained operator is self--adjoint. This can be easily
shown using the following relation
\begin{equation} 
p'_\mu \Gamma^\mu(p)D^j(p,\,p') = -D^j(p,\,p')p_\mu \Gamma^\mu (p')
\label{p Gamma D = - Dp Gamma}
\end{equation}
which is a consequence of
Eq.~(\ref{Gamma D = D Gamma}).

In the case of a scalar operator the values $n$ will be even and
for a pseudoscalar they will be odd.

Let us consider now the 4-vector operator $j_\mu(0)$.
To parametrize the matrix element one needs a set of
quantities of the appropriate tensor dimension. Using
the variables entering the particle state vectors one can
construct one pseudovector $\Gamma^\mu (p')$ and three
independent vectors:  
$$ 
K_\mu = (p - p')_\mu = q_\mu\,,\quad K'_\mu = (p + p')_\mu \,, 
$$
\begin{equation} 
R_\mu = 
\epsilon _{\mu \,\nu \,\lambda\,\rho}\, p^\nu \,p'\,^\lambda\,
\Gamma^\rho (p')\;.
\label{kk'RG}
\end{equation}
Here $\epsilon _{\mu \,\nu \,\lambda\,\rho}$ is a completely
anti-symmetric pseudo-tensor in four dimensional space-time with
$\epsilon _{0\,1\,2\,3}= -1$.

The operator matrix element contains the matrix elements of the
listed quantities multiplied by
$D^j(p,\,p')$ from the left. Each of such products is to be
multiplied by the sum of linearly independent scalars
constructed while obtaining the parameterization
(\ref{DA(0)j= f_n}):
$$
\langle\,\vec p,\,M,\,j,\,m\,|j_\mu(0)|\,\vec p\,',\,M,\,j,\,m'\,\rangle
$$
$$
= \sum_{m''}\,\langle\,m|D^j(p,\,p')|m''\rangle
\langle\,m''|\,F_1\,K'_\mu + F_2\,\Gamma^\mu (p')
$$
\begin{equation}
+ F_3\,R_\mu + F_4\,K_\mu |m'\rangle\;,
\label{<|j|>=F_is}
\end{equation}
where
\begin{equation}
F_i = \sum _{n=0}^{2j}\,f_{in}(Q^2)(ip_\mu\Gamma^\mu(p'))^n\;.
\label{Fi}
\end{equation}

Let us impose some additional conditions on the operator.

1. Let us require the operator to be self--adjoint. One can
check this condition by use of Eq.~(\ref{p Gamma D = - Dp
Gamma}).  In the right--hand side (r.h.s.) of
Eq.~(\ref{<|j|>=F_is}) we need to modify slightly the vector
multiplied by $F_2$.  The new vector is a linear combination of
the 4--vectors entering Eq.~(\ref{<|j|>=F_is}) and has the
following form:  \begin{equation} \Gamma^\mu(p')\;\to\;
\Gamma^\mu(p') - \frac{{K'}\,^\mu}{{K'}^2}\,\left(p_\mu
\Gamma^\mu(p')\right) \;.  \label{G-K} \end{equation} The terms
containing $F_2$ and $F_3$ are modified in the following way:
\begin{equation}
F_i\,A^\mu\;\to\;\frac{1}{2}\left(F_i\,A^\mu + A^\mu\,F_i\right) =
\left\{F_i\,A^\mu\right\}_+\;,\quad i=2,3\;.
\label{+}
\end{equation}
Here $A^\mu$ are the vectors entering Eq.~(\ref{<|j|>=F_is}) and
changed following Eq.~(\ref{G-K}).  The terms containing $F_3$
and $F_4$ have to be multiplied by $i$.

2. It is useful to modify the parametrization so as
to make the vectors -- multipliers of
$F_i$ -- orthogonal to each other.
This results in the following form of the vector multiplied by
$F_2$ (\ref{G-K}):
\begin{equation}
\Gamma^\mu(p') - \left(\frac{{K'}\,^\mu}{{K'}^2} +
\frac{{K}\,^\mu}{{K}^2}\right)\left(p_\mu \Gamma^\mu(p')\right)\;.
\label{G-KKs}
\end{equation}

3. Let us impose the condition of parity conservation.
This condition is satisfied if all the terms in the sum
(\ref{Fi}) contain an even number of pseudovector factors $\Gamma^\mu$.
So, the sums in $F_1$ and $F_4$ are over even
$n:\;2j\;\geq n\;\geq \;0$, in $F_3$ over even $n:\;2j - 1\;\geq
\;n\;\geq \;0$, and in $F_2$ over odd $n:\; 2j -1\;\geq\;n\;>\;0$.

4. Let us impose the conservation condition
$j_\mu K^\mu = j_\mu q^\mu= 0$.
It is easy to see that this condition is satisfied only if
$F_4 = 0$.

So, finally we have
$$
\langle\,\vec p,\,M,\,j,\,m\,|j_\mu(0) |\,\vec p\,',\,M,\,j,\,m'\,\rangle
$$
$$
= \sum_{m''}\,\langle\,m|D^j(p,\,p')|m''\rangle
\langle\,m''|\,F_1K'_\mu + \left\{F_2\left[\Gamma^\mu (p')\right.\right.
$$
\begin{equation}
\left.\left.
- (p_\mu \Gamma^\mu(p'))
\left(\frac {K'_\mu}{{K'}^2}+\frac {K_\mu}{{K}^2}\right )\right]\right\}_+ +
i\left\{F_3\,R_\mu\right\}_+|m'\rangle,
\label{<|j|>=F_i}
\end{equation}

This construction can be used, for example, to obtain the
electromagnetic current matrix element in the case of particle
with spin 1/2.

Now let us list the conditions for the electromagnetic current operator
to be fulfilled in relativistic case
(see Ref.~\cite{KrT02} and the references therein).

(i).{\it Lorentz covariance}:
\begin{equation}
\hat U^{-1}(\Lambda )\hat j^\mu (x)\hat U(\Lambda ) =
\Lambda ^\mu_{\>\nu}\hat j^\nu (\Lambda ^{-1}x)\;.
\label{UjmuU}
\end{equation}
Here $\Lambda $ is a Lorentz--transformation matrix,
$\hat U(\Lambda ) $ is an operator of the unitary
representation of the Lorentz group.\\
(ii).{\it Invariance under translation}:
\begin{equation}
\hat U^{-1}(a)\hat j^\mu(x)\hat U(a) = \hat j^\mu(x-a)\;.
\label{UajmuUa}
\end{equation}
Here $\hat U(a)$  is an operator of the unitary
representation of the translation group.\\
(iii).{\it Current conservation law}:
\begin{equation}
[\,\hat P_\nu\,\hat j^\nu(0)\,] = 0\;.
\label{Pj=0}
\end{equation}
In terms of matrix elements
$\langle\,\hat j^\mu(0)\,\rangle$
the conservation law can be written in the following form:
\begin{equation}
q_\mu\,\langle\,\hat j^\mu(0)\,\rangle  = 0\;.
\label{Qj=0}
\end{equation}
Here $q_\mu$ is four-vector of the momentum transfer.\\
(iv).{\it Current--operator transformations under space--time
reflections}:
$$
\hat U_P\left(\,\hat j^0(x^0\,,\vec x)\,,\hat{\vec j}(x^0\,,\vec x)\right)
\hat U^{-1}_P
$$
$$
=
\left(\,\hat j^0(x^0\,,-\,\vec x)\,,-\,\hat{\vec j}(x^0\,,-\,\vec x)\right)\;,
$$
\begin{equation}
\hat U_R\,\hat j^\mu(x)\,\hat U^{-1}_R = \hat j^\mu(-\,x)\;.
\label{UpUr}
\end{equation}
In Eq.~(\ref{UpUr}) $\hat U_P$ is the unitary operator for the
representation of space reflections and
$\hat U_R$ is the anti-unitary operator of the representation of
space-time reflections
$R = P\,T$.

Our parametrization satisfies all the
conditions listed above.
The parametrization (\ref{<|j|>=F_i})
for the electromagnetic current in the case of particle with
spin 1/2 has the following form:
$$
\langle\,\vec p,\,M,\,\frac{1}{2},\,m\,|\,j_\mu(0)\, |\,\vec
p\,',\,M,\,\frac{1}{2},\,m'\,\rangle
$$
$$
= \sum_{m''}\,\langle\,m|D^{1/2}(p,\,p')|m''\rangle
$$
\begin{equation}
\times
\langle\,m''|\,f_{10}(Q^2)\,K'_\mu +
if_{30}(Q^2)\,R_\mu|m'\rangle\;,
\label{<|j|>=K+R}
\end{equation}
The form factors $f_{10}(Q^2)= f_1(Q^2)$ and $f_{30}(Q^2)=f_2(Q^2)$  are the 
electric and the magnetic form factors of the particle, respectively.
These form factors can be rewritten in terms of standard Sachs
form factors $G_E(Q^2)$ (electric) and $G_M(Q^2)$ (magnetic)
\cite{BaY95}:  
$$ 
f_1(Q^2) = \frac {2M}{\sqrt {4M^2 + Q^2}}\,G_E(Q^2)\;, 
$$ 
\begin{equation} 
f_2(Q^2) = -\frac{4}{M\sqrt {4M^2 + Q^2}}\,G_M(Q^2)\;.  
\label{f=G}
\end{equation}

In analogous way (although a little more cumbersome) one can
obtain the matrix elements of the operators of higher
tensor dimension. This interesting problem, however, is
out of scope of the present paper.

\section{Parametrization of matrix elements of the
two--particle electromagnetic current operator}

To describe the properties of the system of interacting
constituents in our approach it is necessary to have the reduced
matrix elements on Poincar\'e group (the form factors) which
describe the properties of the composite system of free
constituents. In this section we generalize the method of
parametrization of the previous section to the case of such free
systems (see
Refs.~\cite{TrS69,KoT72}, too).

Let us consider a system of two free particles with spins 1/2 and let us
parametrize the matrix element describing the transitions in
this system. Let us construct, for example, the electromagnetic
current operator matrix element. The matrix element can be taken
between the following two--particle state vectors:
\begin{equation}
|\,\vec p_1\,,m_1;\,\vec p_2\,,m_2\,\!\rangle =
|\,\vec p_1\,,m_1\,\!\rangle\otimes |\,\vec p_1\,,m_2\,\rangle\;.
\label{p1m1p2m2}
\end{equation}
Here $\vec p_1\;,\;\vec p_2$ are three--momenta of particles,
$m_1\;,\;m_2$ are spin projections on the $z$ axis.
The one--particle state vectors are normalized by
Eq.~(\ref{normg}).

As well as the basis (\ref{p1m1p2m2}) one can choose another
set of two-particle state vectors
where the motion of the
two-particle center of mass is separated:
$$
|\,\vec P,\;\sqrt {s},\;J,\;l,\;S,\;m_J\,\rangle\;,
$$
$$
\langle\,\vec P,\;\sqrt {s},\;J,\;l,\;S,\;m_J
|\,\vec P\,',\;\sqrt {s'},\;J',\;l',\;S',\;m_{J'}\,\rangle
$$
\begin{equation}
= N_{CG}\,\delta^{(3)}(\vec P - \vec P\,')\delta(
\sqrt{s} - \sqrt{s'})\delta_{JJ'}\delta_{ll'}\delta_{SS'}\delta_{m_Jm_{J'}}\;,
\label{bas-cm}
\end{equation}
$$
N_{CG} = \frac{(2P_0)^2}{8\,k\,\sqrt{s}}\;,\quad
k = \frac{\sqrt{\lambda(s\,,\,M_1^2\,,\,M_2^2)}}{2\,\sqrt{s}}\;,
$$
Here $P_\mu = (p_1 +p_2)_\mu$, $P^2_\mu = s$, $\sqrt {s}$
is the invariant mass of the two-particle system,
$l$ is the orbital angular momentum in the center--of--mass
frame (c.m.), $S$ is the total spin in the c.m., $J$ is the
total angular momentum with the projection $m_J$; $M_1\;,\;M_2 $
are the constituent masses, and 
$\lambda (a,b,c) = a^2 + b^2 + c^2 - 2(ab + bc + ac)$.

The basis (\ref{bas-cm}) is connected with the basis
(\ref{p1m1p2m2})
through the Clebsh--Gordan decomposition
for the Poincar\'e group. Here we write the decomposition in
a little more general form than in
Ref.~\cite{KrT02}:
$$ 
|\,\vec P,\;\sqrt {s},\;J,\;l,\;S,\;m_J\,\rangle 
$$ 
$$ 
= \sum_{m_1\;m_2}\,\int \,\frac {d\vec p_1}{2p_{10}}\,
\frac {d\vec p_2}{2p_{20}}\,|\,\vec p_1\,,m_1;\,\vec p_2\,,m_2\,\rangle 
$$
\begin{equation}
\times \langle\,\vec p_1\,,m_1;\,\vec p_2\,,m_2\,|
\,\vec P,\;\sqrt {s},\;J,\;l,\;S,\;m_J\,\rangle\;,
\label{Klebsh}
\end{equation}
where
$$
\langle\,\vec p_1\,,m_1;\,\vec p_2\,,m_2\,|\,\vec P,\;\sqrt {s},
\;J,\;l,\;S,\;m_J\,\rangle
$$
$$
= \sqrt {2s}[\lambda
(s,\,M_1^2,\,M_2^2)]^{-1/2}\,2P_0\,\delta (P - p_1 - p_2)
$$
$$
\times \sum
\langle\,m_1|\,D^{1/2}(p_1\,,P)\,|\tilde m_1\,\rangle
\langle\,m_2|\,D^{1/2}(p_2\,,P)\,|\tilde m_2\,\rangle
$$
$$
\times
\langle\frac{1}{2}\,\frac{1}{2}\,\tilde m_1\,\tilde m_2\,|S\,m_S\,\rangle
\,Y_{lm_l}(\vartheta\,,\varphi )\,
\langle S\,l\,m_s\,m_l\,|Jm_J\rangle\;.
$$
Here $\vartheta\;,\;\varphi$ are the spherical angles of the
vector $\vec p = (\vec p_1 - \vec p_2)/2$
in the c.m., $Y_{lm_l}(\vartheta,\varphi)$ are
spherical harmonics,
$\langle1/2\,1/2\,\tilde m_1\,\tilde m_2\,|S\,m_S\,\rangle\;,\;
\langle S\,l\,m_s\,m_l\,|Jm_J\rangle$ are Clebsh--Gordan
coefficients for the group $SU(2)$, $D^j$ are
the known rotation matrices to be used for correct relativistic
invariant spin addition
\cite{KoT72}, the summation is over
${\tilde m_1,\tilde m_2,m_l,m_S}$.

The decomposition in spherical harmonics and angular momenta summation
in Eq.~(\ref{Klebsh}) are performed in the c.m. and
the result is shifted to an arbitrary frame by use of
$D$--functions
\cite{KoT72}.

The electromagnetic current matrix element for the system of two
free particles taken in the basis
(\ref{p1m1p2m2}) can be written as a sum of the one--particle
current matrix elements:
$$ 
\langle\vec p_1,m_1;\vec p_2,m_2|j_\mu^{(0)}(0)| 
\vec p\,'_1,m'_1;\vec p\,'_2,m'_2\rangle
$$ 
\begin{equation} 
= \langle\vec p_2,m_2|\vec p\,'_2,m'_2\rangle \langle\vec p_1,m_1|j_{1\mu}(0)|
\vec p\,'_1,m'_1\rangle + (1\leftrightarrow 2)\;.
\label{j=j1+j2}
\end{equation}

Each of the one--particle current matrix elements in
Eq.~(\ref{j=j1+j2}) can be written in terms of form factors as in
Section II. In the case of the particles with spin 1/2 we make
use of Eq.(\ref{<|j|>=K+R}). So, in this case the
electromagnetic properties  of the system are defined by the
form factors $f_{1}\;,\;f_{2}$, given by Eqs.~(\ref{<|j|>=K+R}),
(\ref{f=G}).

Now let us construct the electromagnetic current matrix element
for the system of two free particles in the basis
(\ref{bas-cm}) following the previous Section. Let us consider
first a simple case
$J=J'=S=S'=l=l'=0$. (We omit these variables in the state
vectors.) This set of quantum numbers appears, for
example, in the case of pion. Now there is no pseudovector
$\Gamma^\mu$, but along with the scalar $(P - P')^2 = -\,Q^2$
two additional nontrivial scalars do appear $s' = P'\,^2$ and 
$s = P^2$ -- the invariant mass squares for the free two--particle
system in the initial and in the final states.  So, the form
factors entering the parametrization are functions of the
variables $Q^2\;,\;s\;,\;s'$. The current matrix element is
presented by the linear combination of the four--vectors $P_\mu$
and $P'_\mu$:  
$$ 
\langle\vec P,\sqrt s\mid j_\mu^{(0)}(0) \mid\vec P',\sqrt{s'}\rangle 
$$ 
\begin{equation} 
=  (P_\mu + P'_\mu)\,g_1 (s,Q^2,s') + (P_\mu - P'_\mu)\,g_2 (s,Q^2,s')\;.
\label{<|j|>=Pg1Pg2} 
\end{equation}

Making use of the conservation condition
(\ref{Qj=0})
\begin{equation}
j_\mu^{(0)}(0)(P - P')^\mu = 0\;.
\label{conserv}
\end{equation}
we can write the parametrization in the form:
$$
\langle\vec P,\sqrt s\mid j_\mu^{(0)}(0)\mid \vec P',\sqrt{s'}\rangle
$$
\begin{equation}
=  A_\mu (s,Q^2,s')\;g_0 (s,Q^2,s')\;.
\label{<|j|>=A mu g0}
\end{equation}
Here $g_0 (s,Q^2,s')$ is the reduced matrix element. We will
refer to this invariant as to the free two--particle form
factor. The vector
$A_\mu (s,Q^2,s')$ is defined by the current transformation
properties (the Lorentz covariance and the conservation law):
$$ 
A_\mu (s,Q^2,s') 
$$ 
\begin{equation} 
= (1/Q^2)[(s-s'+Q^2)P_\mu + (s'-s+Q^2) P\,'_\mu]\;.  
\label{Amu}
\end{equation}

The free two--particle form factor can be expressed in terms of
the one--particle form factors (\ref{<|j|>=K+R}) (see
Ref.~\cite{KrT02} for details).  To do this one has to perform in
Eq.~(\ref{<|j|>=A mu g0}) the Clebsh--Gordan decomposition of the
irreducible representation (\ref{bas-cm}) into the direct
product of two irreducible representations (\ref{p1m1p2m2}),
(\ref{Klebsh}) and to take into account
Eqs.~(\ref{<|j|>=K+R}), (\ref{j=j1+j2}).  As the form factors are
invariants one can perform the integration in Eq.~(\ref{<|j|>=A
mu g0}) in the coordinate frame with $\vec P\,' = 0\,$, 
$\vec P = (0,\,0,\,P)$.  The explicit form of $g_0 (s,Q^2,s')$ in the
case of two particles with spin 1/2 and mass $M$ can be found in
Ref.~\cite{KrT02}.

Let us perform the analogous parameterization for the set of
quantum numbers in the basis
(\ref{bas-cm}) in the case $J\;,\;J'\ne$ 0 (see
Ref.~\cite{KoT72}, too).
Finally we will take $J = J'$.

The Lorentz covariant properties of the matrix element are
defined (in analogy with
Eq.~(\ref{kk'RG})) by three 4--vectors and one pseudovector:
$$ 
K'_\mu = (P + P')_\mu\;,\quad K_\mu = (P - P')_\mu \;, 
$$ 
\begin{equation}
R_\mu = \epsilon _{\mu \nu \lambda \rho}P^\nu P'^\lambda
\Gamma^\rho (P')\;, \quad \Gamma_\mu(P')\;.
\label{K'KRG}
\end{equation}
The pseudovector $\Gamma_\mu(P)$
does not enter the parametrization because it can be expressed
through $\Gamma_\mu(P')$ by the equation analogous to
Eq.~(\ref{Gamma D = D Gamma}):
$$
\frac{1}{\sqrt{s}}\,\Gamma(P)_\mu\,D^j(P,P') =
D^j(P,P')\left\{\frac{1}{\sqrt{s'}}\Gamma_\mu(P') \right.  
$$ 
$$
\left.  -\,\frac{1}{\sqrt{ss'}}\cdot\frac{\sqrt{s'}\,P_\mu +
\sqrt{s}\,P_\mu'} {P_\nu P'\,^\nu +
\sqrt{ss'}}\left[P_\nu\Gamma^\nu(P')\right]\right\}\;.  
$$
The set of linearly independent matrices to be used for the
decomposition of the current matrix element is obtained
from the vectors
$P_\mu$ and $\Gamma_\mu(P')$ following Eq.~(\ref{pseud}):
\begin{equation} 
D^J(P,\,P')\,(\,P_\mu \Gamma^\mu(P'))^n\;,\quad n = 0,1,\ldots ,2J\;.
\label{pseud2}
\end{equation}

Using the conditions of self--adjointness, current conservation
(\ref{conserv}), parity conservation and orthogonality of the
parametrization vectors to one another (as in
Eq.~(\ref{<|j|>=F_i})) we obtain (see Eq.~(\ref{+}), too):  
$$
\langle\vec P,\sqrt s,J,l,S,m_J |j_\mu^{(0)}|\vec P',\sqrt{s'}
,J,l',S',{m'_{J}}\rangle
$$
$$
=  \sum _{m''_J}\,\langle m_J|D^J(P\,,P')|m''_J\,\rangle
$$
\begin{equation}
\times\langle\,m''_J|\,\sum ^3_{i=1}\,\left\{F^{ll'SS'}_i\,
A^i_\mu (s,Q^2,s')\right\}_+\,|m'_J\,\rangle\;,
\label{j=FA}
\end{equation}
$$
A^1_\mu =\frac{1}{Q^2}\left[(s-s'+Q^2)P_\mu + (s'-s+Q^2) P\,'_\mu\right]\;,
$$
$$
A^2_\mu =
\frac {1}{\sqrt {s'}}\left\{\,\Gamma_\mu(P') - \frac {1}{2\sqrt s}\left[
-(\sqrt s +\sqrt {s'})\frac {K_\mu}{Q^2}\right.\right.
$$
$$
+ \frac{\sqrt {s'}P_\mu + \sqrt s P'_\mu}{PP' + \sqrt {ss'}}
$$
$$
\left.\left. +\frac{\sqrt{s}-\sqrt{s'}}{\lambda(s,-Q^2,s')}\left[
(\sqrt{s}+\sqrt{s'})^2 + Q^2\right]\,A^1_\mu\right]
\left[P_\lambda \Gamma^\lambda (P')\right]\right\}\,,
$$
\begin{equation}
A^3_\mu = \frac{i}{\sqrt{s'}}\,R_\mu\;.
\label{Ai}
\end{equation}
The quantities
$F^{ll'SS'}_i$ in Eq.~(\ref{j=FA})
are defined by the relations analogous to
Eq.~(\ref{Fi}):
\begin{equation}
F^{ll'SS'}_i = \sum _{n=0}^{2J}\,f^{ll'SS'}_{in}(s\,,\,Q^2\,,\,s')
(iP_\mu \Gamma^\mu(P'))^n\;.
\label{FilS}
\end{equation}
The sum in Eq.~(\ref{FilS}) is taken using the parity
conservation condition as in Eq.~(\ref{<|j|>=F_i}).

Let us remark that the reduced matrix elements -- the invariant
form factors --- now (in contrast with the form factor
(\ref{<|j|>=A mu g0})) depend on the additional invariant
quantities $l,\,l',\,S,\,S'$ that are invariant
degeneration parameters in the basis
(\ref{bas-cm}).

The self--adjointness condition is fulfilled in
Eq.~(\ref{j=FA}) if
\begin{equation}
{f^{ll'SS'}}^*_{in}(s, Q^2, s') = f^{l'lS'S}_{in}(s',Q^2, s)\;.
\label{samosop}
\end{equation}
Here the star means the complex conjugation.

Let us consider especially the case
$M_1=M_2=M\,,\,J=J'=S=S'=$1$\,,\,l=l'=$0.
We will use the reduced matrix elements for this case to
calculate $\rho$ -- meson properties in MIA neglecting the
$D$ -- state contribution (see, e.g. Ref.~\cite{CaG95pl}). In
this case the functions $F^{ll'SS'}_i$ in Eq.~(\ref{FilS}) have
the following form (compare to Eqs.~(\ref{Fi}),
(\ref{<|j|>=F_i})):
$$ F_1 = f_{10}(s,Q^2,s') +
f_{12}(s,Q^2,s')(iP_\nu\Gamma^\nu(P'))^2\;, $$ \begin{equation}
F_2 = f_{21}(s,Q^2,s')\,(i\,P_\nu\Gamma^\nu (P'))\;,\quad
F_3 = f_{30}(s,Q^2,s')\;.
\label{3Fi}
\end{equation}
In equations (\ref{3Fi}) the fixed variables $l,l',S,S'$ are omitted.
Time reflection invariance imposes the following conditions:
\begin{equation}
f^*_{in} = f_{in}\;,\quad i=1,3\;;
\quad f^*_{21} = -\,f_{21}\;.
\label{otrt}
\end{equation}

With Eqs.~(\ref{samosop}) and (\ref{otrt}) we obtain
$$
f_{in}(s, Q^2, s') = f_{in}(s', Q^2, s)\;,
\quad i = 1,3\;;
$$
\begin{equation}
f_{21}(s, Q^2, s') = -\,f_{21}(s', Q^2, s)\;.
\label{f=f^t}
\end{equation}
The relation (\ref{f=f^t}) demonstrates that the form factor
$f_{21}$ appears in the parametrization as a consequence of the
fact that the invariant masses in the initial and the final
states are different. This form factor gives no contribution to
elastic processes, for example, to the electron scattering by
the composite particle, however, it does contribute to radiative
transitions.

Let us rewrite Eq.~(\ref{3Fi}) in terms of standard form factors
instead of $f_{in}$:
$$
F_1 = g_{0C}(s,Q^2,s') + g_{0Q}(s,Q^2,s')
\left\{(iP_\nu \Gamma^\nu(P'))^2 \right. -
$$
$$
\left. - (1/3)\,
\hbox {Sp}(iP_\nu \Gamma^\nu (P'))^2\right\}
\frac{2}{\hbox {Sp}(P_\nu \Gamma^\nu (P'))^2}\;,
$$
$$
F_2 = g_{0MQ}(s,Q^2,s')\,(i\,P_\nu\Gamma^\nu (P'))\;,
$$
\begin{equation}
F_3 = g_{0M}(s,Q^2,s')\;.
\label{3g}
\end{equation}
The scalar factor in the term with $g_{0Q}$
is chosen in such a way as to make it possible to interpret
$g_{0Q}$ as a quadrupole form factor of the system of two free
particles. In other terms
$g_{0C}$ is the charge form factor, $g_{0MQ}= f_{21}$
is the magnetic quadrupole form factor of the second kind, its
classical analog being the so called toroidal
magnetic moment
\cite{DuT90}, $g_{0M} = f_{30}$ is the magnetic form factor of the free two--
particle system.

In the same way as in the case of
Eq.~(\ref{<|j|>=A mu g0}) (see Ref.~\cite{KrT02} for the details)
the free two--particle form factors in
Eq.~(\ref{3g}) can be written in terms of the constituent form
factors. The corresponding equations are rather complicated, so
we present here only the equations that we will use for the
$\rho$--meson electromagnetic structure, that is for the case
when the constituents are the
$u$-- and $\bar{d}$-- quarks (see Appendix 1).

The free two--particle form factors in Eqs.~(\ref{<|j|>=A mu
g0}), (\ref{j=FA}), (\ref{3Fi}), and (\ref{3g}) are to be
considered in the sense of distributions \cite{KrT02}.  For
example, $g_{0}(s\,,Q^2\,,s')$ in Eq.~(\ref{<|j|>=A mu g0}) has
to be interpreted as a Lorentz invariant regular generalized
function on the space of test functions ${\cal S}$({\bf R}$^2$)
\cite{BoL87}.

Let us define the functional giving the regular generalized
function as
$$
\langle\,g_{0} (s,Q^2,s')\,,\varphi(s,s')\rangle
$$
\begin{equation}
= \int\,d\mu(s,s')\,g_{0}(s,Q^2,s')\,\varphi(s,s')\;.
\label{<>}
\end{equation}
Here
$$
d\mu(s,s') = 16\,\theta(s - 4M^2)\,\theta(s' - 4M^2)
$$
\begin{equation}
\times\sqrt[4]{ss'}\,d\mu(s)\,d\mu(s')\;,\quad
d\mu(s) = \frac{1}{4}k\,d\sqrt{s}\;.
\label{dmu}
\end{equation}
The quantity $Q^2$ is a parameter of the generalized function, $M_1=M_2=M$.
The function $\theta(x)$ is the step function.

$\varphi(s\,,s')$ is a function from the space of test functions.
So, for example, the limit as
$Q^2\;\to\;$0 (the static limit) in
$g_0(s\,,Q^2\,,s')$ (\ref{<|j|>=A mu g0})
exists only in the weak sense as the limit of the functional:
\begin{equation}
\lim_{Q^2\to 0}\langle \,g_{0}, \varphi\,\rangle =
\langle(\hbox{e}_q + \hbox{e}_{\bar q})
\delta(\mu(s') - \mu(s)), \varphi\,\rangle\;.
\label{lim Q2=0}
\end{equation}
$\hbox{e}_q$ and $\hbox{e}_{\bar q}$ are the constituent
charges, $\delta$ is the Dirac delta--function.

It is just the limit in the sense
(\ref{lim Q2=0}) that gives the electric charge of the free
two--particle system. The ordinary point--wise
limit of the form factor
$g_{0}$ as $Q^2\to$0 is zero.
The equations analogous to Eq.~(\ref{lim Q2=0})
are valid for the static limits of the free two--particle
form factors in
Eqs.~(\ref{3g}), (A1), (A2) ¨ (A3), too.

Let us note that the conditions imposed on the free
two--particle form factors that follow from the conditions
for the electromagnetic current operator
(\ref{samosop}), and (\ref{f=f^t})
have to be considered in the weak sense, too.

\section{Parameterization of the current operator matrix
elements for systems of two interacting particles}

In this Section we generalize the parameterization method
of the previous sections to the case of composite system with the
structure defined by the interaction of its
constituents.

Let us consider the operator
$j_\mu(0)$ that describes a transition between two states of a
composite two-- constituent system, $j_\mu(0)$ being diagonal in
the total angular momentum. Let us neglect temporarily the
additional conditions of self--adjointness, parity conservation
etc. in the same way as when constructing the matrix
elements (\ref{<|j|>=F_is}) in Section II.  The Wigner--Eckart
decomposition of the matrix element has the form
(\ref{<|j|>=F_is}), (\ref{Fi}).
To emphasize the fact that the particle is composite, let us
rewrite Eqs.~(\ref{<|j|>=F_is}), (\ref{Fi}) using new notations:
$$
\langle\,\vec p_c,\,m_{Jc}\,|j_\mu(0)|\,\vec p_c\,',m'_{Jc}\,\rangle
$$
$$
= \langle\,m_{Jc}|D^{J_c}(p_c,\,p_c')\,
\left[\,F^c_1\,K'_\mu + F^c_2\,\Gamma^\mu (p_c')  \right.
$$
\begin{equation}
\left. + F^c_3\,R_\mu + F^c_4\,K_\mu\right]|m'_{Jc}\rangle\;,
\label{<|jc|>=F_is}
\end{equation}
Here
\begin{equation}
F^c_i = \sum _{n=0}^{2J_c}\,f^c_{in}(Q^2)(ip_{c\mu}\Gamma^\mu(p_c'))^n\;.
\label{Fic}
\end{equation}
In Eqs.~(\ref{<|jc|>=F_is}), and (\ref{Fic}) 
$(p_c - p_c')^2 = -\,Q^2\;,\; p_{c}^2=p'_{c}\,^2=M_c^2\;,\; M_c$ is the mass 
of the composite particle. In the state vector variables spins and
masses are omitted.

In the frame of RHD the form factors of composite systems
$f^c_{in}$ are to be expressed in terms of RHD wave
functions and constituents form factors.

In RHD a state of two particle interacting system is described
by a vector in the direct product of two one--particle Hilbert
spaces (see, e.g., Ref.~\cite{KrT02}). So, the matrix element in
RHD can be decomposed in the basis (\ref{bas-cm}):  
$$
\langle\vec p_c\,,m_{Já}|j_\mu(0)|\vec p_c\,'\,,m'_{Já}\rangle =
\sum\,\int\,\frac{d\vec P\,d\vec P\,'}{N_{CG}\,N_{CG}'}\,
d\sqrt{s}\,d\sqrt{s'}\,
$$
$$
\times\langle\,\vec p_c\,,m_{Jc}|\vec P\,,\sqrt{s}\,,J\,,l\,,S\,,m_J\rangle
$$
$$
\times\langle\vec P\,,\sqrt{s}\,,J\,,l\,,S\,,m_J|j_\mu(0)|
\vec P\,'\,,\sqrt{s'}\,,J'\,,l'\,,S'\,,m_{J'}\rangle
$$
\begin{equation}
\times\langle
\vec P\,'\,,\sqrt{s'},\,J'\,,l'\,,S'\,,m_{J'}|\vec p_c\,'\,,m'_{Jc}\rangle\;.
\label{j=int}
\end{equation}
Here the sum is over variables $J$,$J'$,$l$,$l'$,$S$,$S'$,$m_J$,$m_{J'}$, and
$\langle\vec P\,'\,,\sqrt{s'}\,,J'\,,l'\,,S'\,,m_{J'}|
\vec p_á\,'\,,m'_{Jc}\rangle$ is the wave function in the sense of IF RHD.

Let us write the IF RHD wave function in the form slightly more
general than in Ref.~\cite{KrT02}:
$$
\langle\vec P\,\,,\sqrt{s}\,,J\,,l\,,S\,,m_{J}|\,\vec p_c\,,m_{J_c}\rangle
$$
\begin{equation}
= N_C\,\delta (\vec P\, - \vec p_c)\delta_{J_cJ}\delta_{m_{J_c}m_{J}}
\,\varphi^{J_c}_{lS}(k)\;.
\label{wf}
\end{equation}
$$
N_C = \sqrt{2p_{c0}}\sqrt{\frac{N_{CG}}{4\,k}}\;,
$$
The RHD wave function of constituents relative motion with
fixed total angular momentum is defined as
\begin{equation}
\varphi^{J_c}_{lS}(k(s)) =\sqrt{\sqrt{s}(1 - \eta^2/s^2)}\,u_{lS}(k)\,k\;,
\label{phi(s)}
\end{equation}
and is normalized by the condition
\begin{equation}
\sum_{lS}\int\,u_{lS}^2(k)\,k^2\,dk = 1\;.
\label{norm}
\end{equation}
Here $\eta = M_1^2 - M_2^2\;$,$u_{lS}(k)$  is a model wave
function.

To calculate the form factors in
Eqs.~(\ref{<|jc|>=F_is}), and (\ref{Fic})  let us write the
Wigner--Eckart decomposition on the Poincar\'e group for the
current matrix element in the r.h.s. of Eq.~(\ref{j=int}) should
be written.  However, now there are some difficulties.

In the previous sections we were dealing with the parametrization
of local operator matrix elements in the case when the
transformations of the state vectors and of the operators were
defined by  one and the same representation of the quantum
mechanical Poincar\'e group.
While describing the composite systems in RHD a different
situation can arise when the state vectors and the operator
under consideration are transformed following different
representations of this group.

It is just such a situation takes place in the case of the
matrix element in the r.h.s. of
Eq.~(\ref{j=int}). The operator describes the system of two
interacting particles and transforms following the
representation with Lorentz boosts generators depending on the
interaction \cite{KrT02}.
The state vectors physically describe the system
of two free particles and present the basis of a representation
with interaction--independent generators. So, the Wigner--Eckart
decomposition in the form used in Section II can not be applied
directly to the matrix element in the integrand in the r.h.s. of
Eq.~(\ref{j=int}). This is caused by the fact that it is
impossible to construct 4--vectors describing the matrix element
transformation properties under the action of Lorentz boosts
from the variables entering the state vectors (contrary to the
case of, e.g., Eq.~(\ref{<|jc|>=F_is})).
In fact, the possibility of matrix
element representation in the form (\ref{<|jc|>=F_is})
is based on the following fact. Let us act by Lorentz
transformation on the operator
\begin{equation}
\hat U^{-1}(\Lambda)j^\mu(0)\hat U(\Lambda) = \tilde j^\mu(0)\;.
\label{UjU=tj}
\end{equation}
We obtain the following chain of equalities:
$$
\langle \vec p_c\,,m_{Jc}|\tilde j^\mu(0)|\vec p_c\,'\,,m'_{Jc}\rangle
$$
$$
= \langle \vec p_c\,,m_{Jc}|\hat U^{-1}(\Lambda)j^\mu(0)\hat U(\Lambda)
|\vec p_c\,'\,,m'_{Jc}\rangle
$$
$$
= \sum_{\tilde m_{Jc},\tilde m'_{Jc}}
\langle\,m_{Jc}|[D^{J_c}(R_\Lambda)]^{-1}|\,\tilde m_{Jc}\rangle
$$
\begin{equation}
\times\langle \Lambda \vec p_c\,,\tilde m_{Jc}|j^\mu(0)|
\Lambda \vec p_c\,'\,,\tilde m'_{Jc}\rangle
\langle\,\tilde m'_{Jc}|D^{J_c}(R_\Lambda)|\,m'_{Jc}\rangle.
\label{tj=LpjLp}
\end{equation}
Here $D^{J_c}(R_\Lambda)$
is rotation matrix realizing the angular momentum
transformation under the action of Lorentz transformations.
The equalities (\ref{tj=LpjLp})
show that the transformation properties of the current as a
4--vector (\ref{UjmuU}) can be described using the 4--vectors
of the initial and the final states. This means that
the canonical parameterization is the realization
of
the Wigner--Eckart theorem on the Poincar\'e group.

In the case of the current matrix element in the r.h.s. of
Eq.~(\ref{j=int}) the relations
(\ref{tj=LpjLp}) are not valid and direct application
of the Wigner--Eckart theorem is impossible.

However, it can be shown that for the matrix element in
Eq.~(\ref{j=int}) considered as a generalized function (that is
considered as an object having sense only under integrals and
sums in
Eq.~(\ref{j=int})), the equality (\ref{tj=LpjLp}) is valid in the
weak sense.

Let us consider the matrix element in question as a regular
Lorentz covariant generalized function (see, e.g.,
Ref.~\cite{BoL87}). Using Eq.~(\ref{wf}) let us rewrite
Eq.~(\ref{j=int}) in the following form:
$$
\langle\vec p_c\,,m_{Jc}|j_\mu(0)|\vec p_c\,'\,,m'_{Jc}\rangle =
$$
$$
= \sum_{l,l',S,S'}\int\,\frac{N_c\,N'_c}{N_{CG}\,N_{CG}'}\,
d\sqrt{s}\,d\sqrt{s'}\,\varphi^{Jc}_{lS}(s)\varphi^{Jc}_{l'S'}(s')
$$
\begin{equation}
\times\langle\vec p_c\,,\sqrt{s}\,,J_c\,,l\,,S\,,m_{Jc}|j_\mu(0)|
\vec p_c\,'\,,\sqrt{s'}\,,J_c\,,l'\,,S'\,,m'_{Jc}\rangle\;.
\label{j=int ds}
\end{equation}
Here it is taken into account that the current operator $j_\mu(0)$
is diagonal in total angular momentum of the composite system.

Let us make use of the fact that the set of the states
(\ref{bas-cm}) is complete:
$$
\hat I = \sum\int\,\frac{d\vec P}{N_{CG}}\,d\sqrt{s}\,
$$
\begin{equation}
\times|\vec P\,,\sqrt{s}\,, J\,,l\,,S\,,m_{J}\rangle
\langle\vec P\,,\sqrt{s}\,, J\,,l\,,S\,,m_{J}|\;.
\label{I=compl}
\end{equation}
Here the sum is over all the discrete variables of the basis
(\ref{bas-cm}).

Under the integral the matrix element of the transformed current
satisfies the following chain of equalities (
(\ref{wf}) and  (\ref{I=compl}) are taken into
account):
$$ 
\sum\int\,\frac{N_c\,N_c'}{N_{CG}\,N_{CG}'}\,d\sqrt{s}\,d\sqrt{s'}\,
\varphi^{Jc}_{lS}(s)\varphi^{Jc}_{l'S'}(s')
$$
$$
\times
\langle\vec p_c\,,\sqrt{s}\,,J_c\,,l\,,S\,,m_{Jc}|
\hat U^{-1}(\Lambda)j_\mu(0)\hat U(\Lambda)
$$
$$
\times|\vec p_c\,'\,,\sqrt{s'}\,, J_c\,,l'\,,S'\,,m'_{Jc}\rangle
$$
$$
= \langle\,\vec p_c\,,m_{Jc}|
\hat U^{-1}(\Lambda)\,\hat I\,j_\mu(0)\,\hat I\,\hat U(\Lambda)
|\,\vec p_c\,'\,,m'_{Jc}\rangle
$$
$$
= \sum_{\tilde m_{Jc},\tilde m'_{Jc}}
\langle\,m_{Jc}|[D^{J_c}(R_\Lambda)]^{-1}|\,\tilde m_{Jc}\rangle
$$
$$
\times\langle \Lambda \vec p_c\,,\tilde m_{Jc}|
\hat I\,j^\mu(0)\hat I\,|
\Lambda \vec p_c\,'\,,\tilde m'_{Jc}\rangle
\langle\,\tilde m'_{Jc}|D^{J_c}(R_\Lambda)|\,m'_{Jc}\rangle
$$
$$
= \sum\int\,\frac{N_c\,N_c'}{N_{CG}\,N_{CG}'}\,
d\sqrt{s}\,d\sqrt{s'}\,\varphi^{Jc}_{lS}(s)\varphi^{Jc}_{l'S'}(s')
$$
$$
\times\sum_{\tilde m_{Jc},\tilde m'_{Jc}}
\langle\,m_{Jc}|[D^{J_c}(R_\Lambda)]^{-1}|\,\tilde m_{Jc}\rangle
$$
$$
\times\langle\Lambda\vec p_c\,,\sqrt{s}\,,J_c\,,l\,,S\,,\tilde m_{Jc}|j_\mu(0)
|\Lambda\vec p_c\,'\,,\sqrt{s'}\,,J_c\,,l'\,,S'\,,\tilde m'_{Jc}\rangle
$$
\begin{equation}
\times\langle\,\tilde m'_{Jc}|D^{J_c}(R_\Lambda)|\,m'_{Jc}\rangle\;.
\label{j=int3}
\end{equation}

It is easy to see that under the integral the current matrix
element satisfies the equalities analogous to
Eq.~(\ref{tj=LpjLp}), so now it is possible to use the
parameterization method of the previous sections under the
integral, that is to use the Wigner--Eckart theorem in the weak
sense.

The next step is a parameterization of the matrix element in the
r.h.s. of Eq.~(\ref{j=int ds}).  The r.h.s. can be written as a
functional on the space of test functions of the form (see
Eq.~(\ref{phi(s)}, too)):
\begin{equation}
\psi^{ll'SS'}(s\,,\,s') = u_{lS}(k(s))\,u_{l'S'}(k(s'))\;.
\label{psi(ss')} 
\end{equation}
and Eq.~(\ref{j=int ds}) can be
rewritten as a functional in {\bf R}$^2$ with variables $(s,s')$
(see Eqs.~(\ref{<>})--(\ref{dmu}), too):  
$$ 
\langle\vec p_c\,,m_{Jc}|j_\mu(0)|\vec p_c\,'\,,m'_{Jc}\rangle 
$$ 
$$ 
= \sum_{l,l',S,S'}\int\,d\mu(s,s')\frac{N_c\,N'_c}{N_{CG}\,N_{CG}'}\,
\psi^{ll'SS'}(s,s')
$$
\begin{equation}
\times\langle\vec p_c\,,\sqrt{s}\,,J_c\,,l\,,S\,,m_{Jc}|j_\mu(0)|
\vec p_c\,'\,,\sqrt{s'}\,,J_c\,,l'\,,S'\,,m'_{Jc}\rangle\;.
\label{j=int dmu}
\end{equation}
The measure in the integral (\ref{j=int dmu})
is chosen with the account of the
relativistic density of states, subject to the normalization
(\ref{phi(s)}), (\ref{norm}) (see Eq.~(\ref{dmu}, too)):
$$
d\mu(s,s') = 16\,\theta(s - (M_1+M_2)^2)\,\theta(s' - (M_1+M_2)^2)\,
$$
\begin{equation}
\times\sqrt{\sqrt{s}(1 - \eta^2/s^2)\sqrt{s'}(1 - \eta^2/{s'\,}^2)}\,
d\mu(s)\,d\mu(s')\;.
\label{dmuM1M2}
\end{equation}
$d\mu(s)$ is given by Eq.~(\ref{dmu})).

The sums over discrete invariant variables can be transformed
into integrals by introducing the adequate delta--functions.
Then the obtained expressions are functionals
in {\bf R}$^6$.

The functional in the r.h.s. of
Eq.~(\ref{j=int dmu}) defines a Lorentz covariant generalized
function, generated by the current operator matrix element.
The integral in Eq.~(\ref{j=int dmu})
converges uniformly due to the definition (\ref{psi(ss')}).

Taking into account
Eq.~(\ref{j=int3}) we decompose the matrix element in the r.h.s.
of Eq.~(\ref{j=int dmu}) into the set of linearly independent
scalars entering the r.h.s. of Eq.(\ref{<|jc|>=F_is}):
$$
\frac{N_c\,N'_c}{N_{CG}\,N_{CG}'}\,
$$
$$
\times\langle\vec p_c\,,\sqrt{s},J_c,l,S,m_{Jc}|j_\mu(0)|\vec
p_c\,'\,,\sqrt{s'},J_c,l',S',m'_{Jc}\rangle
$$
$$
= \langle\,m_{Jc}|D^{J_c}(p_c,\,p_c')
\sum_{n=0}^{2J_c}(ip_{c\mu}\Gamma^\mu(p_c'))^n\,
$$
\begin{equation}
\times{\cal A}^{ll'SS'}_{n\mu}(s,Q^2,s')|m'_{Jc}\rangle,
\label{<|jc|>=cA}
\end{equation}
Here ${\cal A}^{ll'SS'}_{n\mu}(s,Q^2,s')$ is a
Lorentz covariant generalized function.

Taking into account
Eq.~(\ref{<|jc|>=cA}) and comparing
the r.h.s. of
Eq.~(\ref{<|jc|>=F_is}) with Eq.~(\ref{j=int dmu}) we obtain:
$$
\sum_{l,l',S,S'}\int\,d\mu(s,s')\,
\psi^{ll'SS'}(s,s')
$$
$$
\times\langle\,m_{Jc}|{\cal A}^{ll'SS'}_{n\mu}(s,Q^2,s')\,|m'_{Jc}\rangle
$$
$$
= \langle\,m_{Jc}|\left[\,f^c_{1n}\,K'_\mu + f^c_{2n}\,\Gamma^\mu (p'_c)
\right.
$$
\begin{equation}
\left.
+ f^c_{3n}\,R_\mu + f^c_{4n}\,K_\mu\right]|m'_{Jc}\rangle\;.
\label{c=c}
\end{equation}

All the form factors in the r.h.s. of
Eq.~(\ref{c=c}) are nonzero if the generalized function ${\cal
A}$ contains parts that are diagonal (${\cal A}_1$) and
non-diagonal (${\cal A}_2$) in $m_{Jc}\,,\,m'_{Jc}$:
\begin{equation}
{\cal A}^{ll'SS'}_{n\mu}(s,Q^2,s') = {\cal A}^{ll'SS'}_{1n\mu}(s,Q^2,s') +
{\cal A}^{ll'SS'}_{2n\mu}(s,Q^2,s')\;.
\label{cA=cA1+cA2}
\end{equation}

For the diagonal part we have from
Eq.~(\ref{c=c}):
$$
\sum_{l,l'S,S'}\int\,d\mu(s,s')\,
\psi^{ll'SS'}(s,s')
$$
$$
\times\langle\,m_{Jc}|{\cal A}^{ll'SS'}_{1n\mu}(s,Q^2,s')\,|m_{Jc}\rangle
$$
\begin{equation}
= \langle\,m_{Jc}|\left[\,f^c_{1n}[\psi]\,K'_\mu +
f^c_{4n}[\psi]\,K_\mu\right]|m_{Jc}\rangle\;.
\label{cd=cd}
\end{equation}
The notation $f^c_{in}[\psi]$ in the r.h.s. emphasizes the fact
that form factors of composite systems are
functionals on the wave functions of the intrinsic motion
and so, due to Eq.~(\ref{psi(ss')}), on the test functions.

Let the equality
(\ref{cd=cd}) be valid for any test function
$\psi^{ll'SS'}(s,s')$.
When the test functions (the intrinsic motion wave functions)
are changed the vectors in the r.h.s. are not changed because
according to the essence of the parametrization
(\ref{<|jc|>=F_is}) they do not depend on the model for the
particle intrinsic structure. So, when the test functions are
varied the vector of the r.h.s. of Eq.~(\ref{cd=cd})
remains in the hyperplace defined by the vectors
$K_\mu\,,\,K'_\mu$.

When test functions are varied arbitrarily the vector in the
left--hand side (l.h.s.) of Eq.~(\ref{cd=cd}) can take, in
general, an arbitrary direction. So, the requirement of the
validity of Eq.~(\ref{cd=cd}) in the whole space of our test
functions is that the l.h.s. generalized function should have
the form:  
$$ 
{\cal A}^{ll'SS'}_{1n\mu}(s,Q^2,s') = K'_\mu\,G^{ll'SS'}_{1n}(s,Q^2,s') 
$$ 
\begin{equation} 
+ K_\mu\,G^{ll'SS'}_{4n}(s,Q^2,s')\;.  
\label{cA1} 
\end{equation}
Here $G^{ll'SS'}_{in}(s,Q^2,s')\;,\; i=1,4$
are Lorentz invariant generalized functions. Substituting
Eq.~(\ref{cA1}) in Eq.~(\ref{cd=cd}) and taking into account
Eqs.~(\ref{phi(s)}) and (\ref{dmuM1M2})
we obtain the following integral representations:
$$
f^c_{in}(Q^2) =
\sum_{l,l',S,S'}\int_{M_1+M_2}^{\infty}\,d\sqrt{s}\,d\sqrt{s'}\,
$$
\begin{equation}
\times
\varphi^{Jc}_{lS}(s)\,G^{ll'SS'}_{in}(s,Q^2,s')\varphi^{Jc}_{l'S'}(s')\;.
\label{intrep}
\end{equation}
for $i = 1,4$.
In the case of matrix element in
Eq.~(\ref{c=c}) non-diagonal in $m_{Jc}\,,\,m'_{Jc}$ we can
proceed in an analogous way and obtain an analogous integral
representations for $f^c_{in}(Q^2)\;,\;i=2,3$.

So, the matrix element in the r.h.s. of
Eq.~(\ref{j=int dmu}) considered as Lorentz covariant
generalized function can be written as the following
decomposition of the type of Wigner--Eckart decomposition:  
$$
\langle\vec p_c\,,\sqrt{s}\,,J_c\,,l\,,S\,,m_{Jc}|j_\mu(0)|
\vec p_c\,'\,,\sqrt{s'}\,,J_c\,,l'\,,S'\,,m'_{Jc}\rangle
$$
$$
= \frac{N_{CG}\,N_{CG}'}{N_c\,N'_c}\,
\langle\,m_{Jc}|D^{J_c}(p_c,\,p_c')\left[\,{\cal F}_{1}\,K'_\mu +
{\cal F}_{2}\,\Gamma^\mu (p'_c) \right.
$$
\begin{equation}
\left.
+ {\cal F}_{3}\,R_\mu + {\cal F}_{4}\,K_\mu\right]|m'_{Jc}\rangle\;.
\label{fin}
\end{equation}
\begin{equation}
{\cal F}_i = \sum _{n=0}^{2J_c}\,G^{ll'SS'}_{in}(s,Q^2,s')
(ip_{c\mu}\Gamma^\mu(p_c'))^n\;,
\label{cfic}
\end{equation}
with the constraint (\ref{intrep}).

In Eqs.~(\ref{fin}), and (\ref{cfic})
the form factors contain all the information about the physics
of the transition described by the operator
$j_\mu(0)$. They are connected with the composite particle form
factors (\ref{<|jc|>=F_is}), and (\ref{Fic})
through Eq.~(\ref{intrep}). In particular, physical
approximations are formulated in our approach in terms of form
factors $G^{ll'SS'}_{in}(s,Q^2,s')$ (see Ref.~\cite{KrT02} for
details).  The matrix element transformation properties are
given by the 4--vectors in the r.h.s. of Eq.~(\ref{fin}).

It is worth to emphasize that it is necessary to consider the
composite system form factors as the functionals generated
by the Lorentz invariant generalized functions
$G^{ll'SS'}_{in}(s,Q^2,s')$.

Now let us impose the additional conditions
on the matrix
elements in Eqs.~(\ref{<|jc|>=F_is}), and (\ref{fin})
in the same way as we did in Sec.II in
Eqs.~(\ref{G-K})--(\ref{<|j|>=F_i}) and in Sec.III in
Eqs.~(\ref{samosop}), (\ref{otrt}), and (\ref{f=f^t}).  The
r.h.s.  of equalities (\ref{<|jc|>=F_is}) and  (\ref{fin})
contain the same 4--vectors and the same sets of
Lorentz scalars (\ref{Fic}) and (\ref{cfic}), so, to take into
account the additional conditions it is necessary to redefine
these 4--vectors according to
Eqs.~(\ref{G-K})--(\ref{<|j|>=F_i}).  For example, the
conservation law gives ${\cal F}_{4}$=0.  It is easy to see that
for the redefined form factors the equality (\ref{intrep})
remains valid.

Let us write the parameterization
(\ref{fin}), (\ref{cfic}) for the particular case of composite particle
electromagnetic current with quantum numbers
$J=J'=S=S'=1$, which is realized, for example, in the case of
deuteron. Separating the quadrupole form factor in analogy
with Eq.~(\ref{3g}) and using
Eqs.~(\ref{fin}), and (\ref{cfic}) we obtain the following form:
$$
\langle\vec p_c\,,\sqrt{s}\,,J_c\,,l\,,S\,,m_{Jc}|j_\mu(0)|
\vec p_c\,'\,,\sqrt{s'}\,,J_c\,,l'\,,S'\,,m'_{Jc}\rangle
$$
\begin{equation}
= \frac{N_{CG}\,N_{CG}'}{N_c\,N'_c}\,\langle\,m_{Jc}|\,D^{1}(p_c\,,p'_c)\,
\left[
\tilde{\cal F}_1\,K'_\mu
+ \frac{i}{M_c}\tilde{\cal F}_3\,R_\mu\right]|m'_{Jc}\rangle\;.
\label{J=1}
\end{equation}
$$
\tilde{\cal F}_1 = \tilde G^{ll'}_{10}(s,Q^2,s') +
\tilde G^{ll'}_{12}(s,Q^2,s')\left\{[i{p_c}_\nu\,\Gamma^\nu(p'_c)]^2
\right.
$$
$$
\left. - \frac{1}{3}\,\hbox{Sp}[i{p_c}_\nu\,\Gamma^\nu(p'_c)]^2\right\}
\frac{2}{\hbox{Sp}[{p_c}_\nu\,\Gamma^\nu(p'_c)]^2}\;,
$$
\begin{equation}
\tilde{\cal F}_3 = \tilde G^{ll'}_{30}(s,Q^2,s')\;.
\label{FJ=1}
\end{equation}
We have taken into account that the equation
$\tilde G^{ll'}_{21}(s,Q^2,s') =$ 0 is valid in weak
sense.
The calculation of the form factors in Eqs. (\ref{J=1}),
(\ref{FJ=1}) in MIA will be discussed in the following Section.

\section{Electromagnetic current matrix element for composite systems with\\
$J$ =1}

In this section we make use of the formalism developed in the
previous sections to describe composite systems with $J
= J' = S = S'=$1. The case of zero total angular momentum and
zero total spin was considered in detail in
Ref.~\cite{KrT02}.

If we use the form
(\ref{<|jc|>=F_is}) and impose the additional conditions in
analogy with Eqs. (\ref{J=1}), and (\ref{FJ=1}), then the
parameterization (\ref{<|jc|>=F_is}) takes the following form:
$$
\langle\vec p_c\,,m_{Jc}|j_\mu(0)|\vec p_c\,'\,,m'_{Jc}\rangle
$$
\begin{equation}
= \langle\,m_{Jc}|\,D^{1}(p_c\,,p'_c)\, \sum_{i=1,3}\,
\tilde{F}\,^c_i\,\tilde A^i_\mu\,|m'_{Jc}\rangle\;.
\label{jc=FA}
\end{equation}
Here
$$
\tilde{F}\,^c_1 = \tilde f^c_{10}(Q^2) + \tilde
f^c_{12}(Q^2)\left\{[i{p_c}_\nu\,\Gamma^\nu(p'_c)]^2 \right.
$$
$$
\left. - \frac{1}{3}\,\hbox{Sp}[i{p_c}_\nu\,\Gamma^\nu(p'_c)]^2\right\}
\frac{2}{\hbox{Sp}[{p_c}_\nu\,\Gamma^\nu(p'_c)]^2}\;,
$$
\begin{equation}
\tilde{F}\,^c_3 = \tilde f^c_{30}(Q^2)\;.
\label{FicAic}
\end{equation}
$$
\tilde A^1_\mu = (p_c + p'_c)_\mu\;,\quad
\tilde A^3_\mu = \frac{i}{M_c}
\varepsilon_
{\mu\nu\lambda\sigma}\,p_c^\nu\,p'_c\,^\lambda\,\Gamma^\sigma(p'_c)\;.
$$
The form factors are redefined here (compare to Eq.~(\ref{Fic}))
so as to have the meaning of the charge, quadrupole and magnetic
form factors (see Eq.~(\ref{3g}), too).

It should be mentioned that formally
Eqs.~(\ref{jc=FA}), and (\ref{FicAic}) can be obtained from
Eqs.~(\ref{j=FA}) -- (\ref{3g}) for the current matrix
element of the free system if 
$P_\mu = p_{c\mu}$,$\;P_\mu' = p'_{c\mu}$.

The representation (\ref{jc=FA}), and (\ref{FicAic})
of the matrix element satisfies all the conditions on the
composite system electromagnetic current \cite{KrT02} (see Sec.II).

Our form factors in Eq.~(\ref{FicAic}) can be written in terms
of standard Sachs form factors for the system with the total
angular momentum one.  To do this let us write the
parameterization of the electromagnetic current matrix element
in the Breit frame (see, e.g., Ref.~\cite{ArC80}):  
$$
\langle\vec p_c\,,m_{Jc}|j_\mu(0)|\vec p_c\,'\,,m'_{Jc}\rangle =
G^\mu(Q^2)\;,
$$
$$
G^0(Q^2) = 2p_{c0}\left\{(\vec\xi\,'\vec\xi\,^*)\,G_C(Q^2) \right.
$$
\begin{equation}
\left.
+ \left[(\vec\xi\,^*\vec Q)(\vec\xi\,'\vec Q) - \frac{1}{3}Q^2
(\vec\xi\,'\vec\xi\,^*)\right]\,G_Q(Q^2)\frac{1}{2M_c^2}\right\}\;,
\label{Gi}
\end{equation}
$$
\vec G(Q^2) = \frac{p_{c0}}{M_c}\left[\vec\xi'\,(\vec\xi\,^*\vec Q) -
\vec\xi\,^*(\vec\xi\,'\vec Q)\right]\,G_M(Q^2)\;.
$$
Here $G_C\;,\;G_Q\;,\;G_M$  are the charge, quadrupole and
magnetic form factors, respectively.

The polarization vector in the Breit frame has the
following form:
$$
\xi^\mu(\pm 1) = \frac{1}{\sqrt{2}}(0\;,\;\mp 1\;,\;-\,i\;,\;0)\;,
$$
\begin{equation}
\xi^\mu(0) = (0\;,\;0\;,\;0\;,\;1)\;.
\label{polyar}
\end{equation}
The variables in $\xi$ are total angular momentum
projections.

In the Breit frame:
$$
q^\mu = (p_c - p'_c)^\mu = (0\;,\;\vec Q)\;,
$$
\begin{equation}
p_c^\mu = (p_{c0}\;,\;\frac{1}{2}\vec Q)\;,\quad
p'_c\,^\mu = (p_{c0}\;,\;-\frac{1}{2}\vec Q)\;,
\label{Breit}
\end{equation}
$$
p_{c0} = \sqrt{M_c^2 + \frac{1}{4}Q^2}\;,\quad
\vec Q = (0\;,\;0\;,\;Q)\;.
$$
Comparing Eq.~(\ref{jc=FA}) with Eq.~(\ref{Gi}) and taking into
account the fact that in the Breit system 
$D^1_{m_J\,m''_J}(p_c\,,p_c') = \delta_{m_J\,m''_J}\;.$ 
we have:  
$$ 
G_C(Q^2) = \tilde f^c_{10}(Q^2)\;,\quad G_Q(Q^2) =
\frac{2\,M_c^2}{Q^2}\,\tilde f^c_{12}(Q^2)\;, 
$$
\begin{equation} 
G_M(Q^2) = -\,M_c\,\tilde f^c_{30}(Q^2)\;.
\label{GF} 
\end{equation}

The integral representations for the composite system form factors in
Eq.~(\ref{jc=FA}) are obtained in analogy with
Eq.~(\ref{intrep}) from Eqs.~(\ref{j=int ds}),
(\ref{J=1})--(\ref{FicAic}):
\begin{equation} 
\tilde f^c_{in}(Q^2) = \sum_{l,l'}\int\,d\sqrt{s}\,d\sqrt{s'}\,
\varphi^l(s)\,\tilde G^{ll'}_{in}(s,Q^2,s')\varphi^{l'}(s')\;.
\label{intrepJ1}
\end{equation}
In Eq.~(\ref{intrepJ1}) the variables $S=S'=$1 are omitted

Let us pay a special attention to the following points.

1. The chain of equalities
(\ref{jc=FA})--(\ref{intrepJ1}) guarantees that the general
conditions
 (\ref{UjmuU})--(\ref{UpUr})
for the electromagnetic current operator
are fulfilled on each step of the
calculation.

2.  The result for the form factors (\ref{intrepJ1}) does not
depend on the actual values of $m_J\;,\;m_J'$ entering the
l.h.s. of Eq.  (\ref{jc=FA}).  In the standard light--front
dynamics approach such an ambiguity does exist
\cite{Kei94,CaG95pl}.  

3. As compared with light front dynamics
\cite{ChC88,ChC88pl} we have not used any particular ("good")
current components in the calculation of form factors.

Using the equality (\ref{GF})
we obtain three scalar equalities for the standard Sachs form
factors of composite system with total angular momentum one
and total spin one:
$$ 
G_c(Q^2) =
\sum_{l,l'}\int\,d\sqrt{s}\,d\sqrt{s'}\, \varphi^l(s)\,\tilde
G^{ll'}_{10}(s\,,Q^2\,,s')\, \varphi^{l'}(s')\;.
$$
$$
G_Q(Q^2)
$$
\begin{equation}
= \frac{2\,M_c^2}{Q^2}\,\sum_{l,l'}\int\,d\sqrt{s}\,d\sqrt{s'}\,
\varphi^l(s)\,\tilde G^{ll'}_{12}(s\,,Q^2\,,s')\,
\varphi^{l'}(s')\;.
\label{Gq}
\end{equation}
$$
G_M(Q^2)
$$
$$
= -\,M_c\,\sum_{l,l'}\int\,d\sqrt{s}\,d\sqrt{s'}\,
\varphi^l(s)\,\tilde G^{ll'}_{30}(s\,,Q^2\,,s')\,
\varphi^{l'}(s')\;.
$$

The next case of our consideration is the case of $\rho$ meson.
In our composite quark model the $u$- and $\bar d$-- quarks are
in the $S$ state of the relative motion, that is $l = l' =$ 0.
So there is no summation in Eq.~(\ref{Gq}).

Let us use for (\ref{Gq}) the modified impulse approximation
formulated in terms of form factors
$\tilde G^{ll'}_{iq}(s,Q^2,s')$.
The physical meaning of this approximation is considered in
detail in Ref.~\cite{KrT02}. In the frame of MIA
the invariant form factors $\tilde G^{ll'}_{iq}(s,Q^2,s')$
in (\ref{Gq}) are changed by the free two--particle invariant
form factors given by
Eqs.~(\ref{j=FA}), (\ref{3g}), and (A1)--(A3)
and describing the electromagnetic properties of the system of
two free particles. So, the equations to be used for the
calculation of the $\rho$--meson properties in MIA are the following:
$$
G_c(Q^2) = \int\,d\sqrt{s}\,d\sqrt{s'}\,
\varphi(s)\,g_{0C}(s\,,Q^2\,,s')\, \varphi(s')\;,
$$
\begin{equation}
G_Q(Q^2) =
\frac{2\,M_c^2}{Q^2}\,\int\,d\sqrt{s}\,d\sqrt{s'}\,
\varphi(s)\,g_{0Q}(s\,,Q^2\,,s')\,\varphi(s')\;,
\label{GqGRIP}
\end{equation}
$$
G_M(Q^2) =-\,M_c\,\int\,d\sqrt{s}\,d\sqrt{s'}\,
\varphi(s)\,g_{0M}(s\,,Q^2\,,s')\,
\varphi(s')\;.
$$

It should be noted that
Eq.~(\ref{GqGRIP}) can be obtained formally by changing in the
r.h.s of Eq.~(\ref{j=int}) the current matrix element with
interaction into the current of the free two--particle system
(\ref{j=FA}), (\ref{Ai}), (\ref{3g}) along with the following
change in the covariant part
(\ref{j=FA}):
$$
D^J(P,P')\;,\;A^{i}_\mu\;\to\;
D^J(P,P')\;,\;A^{i}_\mu\left|_{P=p_c\;,P'=p'_c}\right.\;,
$$
\begin{equation}
i=1,2,3
\label{A=P+P'}
\end{equation}
So, in our approach the current for the interacting system (to
be precise, its matrix element) is obtained from the free
two--particle current by including the interaction in the
covariant part of
$D^J\;,\;A^{i}_\mu$ in Eq.~(\ref{j=FA}).
In Eq.~(\ref{A=P+P'}) the components of vectors $p_á\;,\;p'_á$
differ from vectors $P\;,\;P'$ of the free two--particle
system in Eqs.~(\ref{j=FA}), and (\ref{Ai})
as to take the interaction into
account. So, our approach does not contradict to the general
statement \cite{ChC88} saying that the current of an interacting system
should be interaction dependent for the conditions of Lorentz
covariance and the current conservation law be fulfilled.

The detailed discussion of
Lorentz--covariance condition, conservation laws, cluster
separability and of the charge nonrenormalizability,
as well as of the dynamical content of MIA, can be found in
Ref.~\cite{KrT02}.

\section{$\rho$ -- meson electromagnetic structure}

Electroweak properties of composite hadron systems were
described in the RHD approach in a number of papers.
The most popular approach in the frame of RHD is the
light--front dynamics
\cite{ChC88,Kei94,CaG95pl,GrK84,ChC88pl,Ter76,Sch94}.
Recently some calculations in the frame of instant form
\cite{KrT93,ChY99}  and point form dynamics
\cite{AlK98} have appeared. The
$\rho$ -- meson electromagnetic structure was calculated in
Refs.~\cite{Kei94,CaG95pl} in the light--front dynamics approach.

In this section we make use of the results
obtained in the previous sections to
calculation  the
$\rho$ -- meson electromagnetic properties.

The $\rho$ -- meson electromagnetic form factors are calculated
using Eqs.~(\ref{GqGRIP}) in MIA. The wave functions are taken according
to Eqs.~(\ref{phi(s)}),  and (\ref{norm}). We suppose that
quarks are in the $S$ state of relative motion.

For the description of the relative motion of quarks (as in
Ref.~\cite{KrT02} in the case of pion) in
Eq.~(\ref{phi(s)}) the following phenomenological wave functions
are used:

1. A Gaussian or harmonic oscillator (HO) wave function
(see, e.g., Ref.~\cite{ChC88pl})
\begin{equation}
u(k)= N_{HO}\,
\hbox{exp}\left(-{k^2}/{2\,b^2}\right),
\label{HO-wf}
\end{equation}

2. A power-law (PL)  wave function \cite{Sch94}:
\begin{equation}
u(k) =N_{PL}\,{(k^2/b^2 +
1)^{-n}}\;,\quad n = 2\;,3\;.
\label{PL-wf}
\end{equation}

3. The wave function with linear confinement from Ref.~\cite{Tez91}:
\begin{equation} u(r) = N_T \,\exp(-\alpha
r^{3/2} - \beta r)\;, \label{Tez91-wf} \end{equation} $$ \alpha
= \frac{2}{3}\sqrt{2\,M_r\,a}\;,\quad \beta = {M_r}\,b\;.  $$
Here $a\;,b$ are the parameters of linear and Coulomb parts of
potential respectively, $M_r$ is the reduced mass of the
two--particle system.

For Sachs form factors of quarks
(see (A1)--(A3)) we have \cite{CaG96}:
\begin{equation} 
G^{q}_{E}(Q^2) = e_q\,f(Q^2)\;,\quad
G^{q}_{M}(Q^2) = (e_q + \kappa_q)\,f_q(Q^2)\;, 
\label{q ff}
\end{equation} 
where $e_q$ is the quark charge and $\kappa_q$ is
the quark anomalous magnetic moment.  However, for $f_q(Q^2)$
the form of Ref.~\cite{KrT98,KrT99} is used instead of Ref.~\cite{CaG96}:
\begin{equation} 
f_q(Q^2) = \frac{1}{1 + \ln(1+\langle r^2_q\rangle Q^2/6)}\;.  
\label{f_qour} 
\end{equation}
Here $\langle r^2_q\rangle$ is the mean square radius (MSR) of constituent
quark.

The details describing the cause of the choice (\ref{f_qour})
for the function $f_q(Q^2)$
can be found in Ref.~\cite{KrT98} (see also Ref.~\cite{KrT02})
and are based on the fact that this form gives the asymptotics
of the pion form factor as $Q^2\to\infty$ that coincides with
the QCD asymptotics \cite{MaM73}.

So, for the calculations we use a standard set of parameters of
CQM.
The structure of the constituent quark is described by the following
parameters:
$M_u = M_{\bar d} = M$ is the constituent quark mass,
$\kappa_u\;,\;\kappa_{\bar d}$ is the constituent quarks
anomalous magnetic moments,
$\langle\,r^2_u\rangle = \langle\,r^2_{\bar d}\rangle = \langle\,r^2_q\rangle$ 
is the quark MSR.
The interaction of quarks in
$\rho$ meson is characterized by wave functions
(\ref{HO-wf}) -- (\ref{Tez91-wf}) with the parameters
$b\;,\;\alpha\;,\;\beta$.

The parameters were fixed in our calculation as follows. We use
$M$=0.25 GeV \cite{KrT01} for the quark mass.
The quark anomalous magnetic moments enter the
equations through the sum ($\kappa_u + \kappa_{\bar d}$) and we take
$\kappa_u + \kappa_{\bar d}$ = 0.09 in natural units following
\cite{Ger95}.

For the quark MSR we use the relation
$\langle r^2_q\rangle \simeq 0.3 /M^2$
\cite{CaG96,VoL90}.

The parameter $b$ in the Coulomb part of the
potential in Eq.~(\ref{Tez91-wf}) is
$b$ = 0.7867.
This gives the value $\alpha_S$ = 0.59 for systems of light quarks
\cite{LuS91}.

We choose the parameters $b$ in Eqs. (\ref{HO-wf}) and
(\ref{PL-wf}) and $a$ in Eq. (\ref{Tez91-wf}) so as to fit the
MSR of $\rho$ meson.  The $\rho$ -- meson MSR is given by
the equation $\langle r^2_\rho\rangle - \langle r^2_\pi\rangle$
= 0.11$\pm$0.06 fm$^2$ \cite{CaG96,VoL90}.  For the pion MSR the
experimental data \cite{Ame84} is taken:  
$\langle r^2_\pi\rangle^{1/2}$ = 0.657$\pm$0.012 fm.  We use the following
relation:
\begin{equation}
\langle\,r^2_\rho\rangle = -6\,G'_{C}(0)\;.
\label{MSR}
\end{equation}

The $\rho$ -- meson mass in Eq.~(\ref{GqGRIP}) is taken from
Ref.~\cite{Abe99}:  $M_\rho$ = 763.0$\pm$1.3 MeV .

The magnetic $\mu_\rho$ and the quadrupole $Q_\rho$
moments of $\rho$ meson were calculated using the relations
\cite{ArC80}:
\begin{equation} 
G_M(0) = \frac{M_\rho}{M}\,\mu_\rho\;,\quad G_Q(0) = M_\rho^2\,Q_\rho\;.
\label{stat}
\end{equation}
The static limit in Eqs.~(\ref{GqGRIP}) gives the following
relativistic expressions for moments:
$$
\mu_\rho = \frac{1}{2}\,\int_{2M}^\infty\,d\sqrt{s}\,\frac{\varphi^2(s)}
{\sqrt{s - 4\,M^2}}\,\left\{1 - L(s) \right.
$$
\begin{equation}
\left.
+ (\kappa_u + \kappa_{\bar d})
\left[1 - \frac{1}{2}\,L(s)\right]\right\}\;,
\label{mu}
\end{equation}
$$
Q_\rho = -\,\frac{1}{4\,M}\,\int_{2M}^\infty\,d\sqrt{s}\,\frac{\varphi^2(s)}
{\sqrt{s(s - 4\,M^2)}}\,\left[\frac{M}{\sqrt{s} + 2\,M} \right.
$$
\begin{equation}
\left.
+ \kappa_u + \kappa_{\bar d}\right]
\,{L(s)}\;,
\label{Q}
\end{equation}
$$
L(s) = \frac{2\,M^2}{\sqrt{s - 4\,M^2}\,(\sqrt{s} + 2\,M)}\,\left[
\frac{1}{2\,M^2}\sqrt{s\,(s - 4\,M^2)} \right.
$$
$$
\left.
+ \ln\,
\frac{\sqrt{s} - \sqrt{s - 4\,M^2}}{\sqrt{s} + \sqrt{s - 4\,M^2}}\right]\;.
$$
Let us note that the nonzero
$\rho$ -- meson quadrupole moment appears due to the
relativistic effect of Wigner spin rotation of quarks only
(see (A2)).  So, measuring of the quadrupole moment
can be a test of the relativistic invariance in the confinement
region.

The Wigner rotation contributes to the magnetic moment, too.
Without spin rotation
($\omega_1 = \omega_2$ = 0 in (A3)) we have for the
magnetic moment:
$$ 
\tilde\mu_\rho = \frac{1}{2}(1 + \kappa_u + \kappa_{\bar d})\, 
$$ 
\begin{equation}
\times\int_{2M}^\infty\,d\sqrt{s}\, \frac{\varphi^2(s)} {\sqrt{s
- 4\,M^2}}\,\left\{1 - \frac{1}{2}\,L(s)\right\}\;,
\label{muwsr}
\end{equation}

To estimate the contribution of the relativistic effects to the
$\rho$ -- meson electromagnetic structure the
non-relativistic calculation of electromagnetic form factors and
static moments was performed. The non-relativistic limit of
equations (\ref{GqGRIP}) gives the following forms of the
form factors:  
$$ 
G_C(Q^2) = \left(G^u_E(Q^2)+G^{\bar d}_E(Q^2)\right)\,I(Q^2)\;,\quad 
G_Q(Q^2) = 0\;, 
$$
\begin{equation}
G_M(Q^2) = \frac{M_\rho}{M}\left(G^u_M(Q^2)+G^{\bar d} _M(Q^2)\right)\,
I(Q^2)\;,
\label{GNR}
\end{equation}
$$
I(Q^2) = \int\,k^2\,dk\,k'\,^2\,dk'\,u(k)\,g_{NR}(k,Q^2,k')u(k')\;.
$$
$$
g_{NR}(k,Q^2,k') = \frac{1}{k\,k'\,Q}\,
$$
$$
\times\left[\theta\left(k' - \left|k - \frac{Q}{2}\right|\right) -
\theta\left(k' - k - \frac{Q}{2}\right)\right]\;.
$$

Let us note that the obtained result
(\ref{GNR}) coincides with the one derived in standard non-relativistic
calculations for composite system form factors in the impulse approximation
\cite{BrJ76}. So, Eq.~(\ref{GqGRIP}) can be considered as a
relativistic generalization of the equations of
Ref.~\cite{BrJ76}.

\begin{table}
\caption{The $\rho$ -- meson statical moments obtained
with the different model wave functions
\protect(\ref{HO-wf}) -- \protect(\ref{Tez91-wf}).
$\langle\,r^2_{NR}\rangle$ is
nonrelativistic MSR, $\langle\,\tilde
r^2\rangle$ is relativistic MSR without spin rotation
contribution, $\mu_\rho$ is relativistic magnetic moment
(\ref{mu}), $\tilde\mu_\rho$ is relativistic magnetic moment
without spin rotation (\ref{muwsr}),
$Q_\rho$ is quadrupole moment (\ref{Q}).
The wave functions parameters are obtained from the fitting of
$\rho$ -- meson MSR obtained through the relativistic
calculation with spin rotation,
$\langle r^2_\rho\rangle - \langle
r^2_\pi\rangle$ = 0.11$\pm$0.06 fm$^2$ \protect\cite{CaG96,VoL90}. The pion
MSR was
taken from the experimental data
\protect\cite{Ame84}:  $\langle r^2_\pi\rangle^{1/2}$ = 0.657$\pm$0.012 fm.
The magnetic moments are in natural units, the quadrupole
moments and MSR are in fm$^2$.
The parameters $b$ in \protect(\ref{HO-wf}) and \protect(\ref{PL-wf}) --
are in GeV and $a$ in \protect(\ref{Tez91-wf}) -- in GeV$^2$.
The values of other parameters are given in the text.
The number of the significant digits is chosen
so as to demonstrate the extent of model dependence
of calculations.}
\begin{tabular}{ccccccc}
~~Wave~~       &       &       &       &       &       &       \\
~~functions~~  &$b\;,a$&$\langle\,r^2_{NR}\rangle$
	       &$\langle\,\tilde r^2\rangle$
	       &$\mu_\rho$&$\tilde \mu_\rho$&$Q_\rho$\\
\noalign{\smallskip}\hline\noalign{\smallskip}
(\ref{HO-wf})    & 0.231 & 0.275 & 0.731 & 0.852 & 0.966 & -0.0065\\
(\ref{PL-wf}) n=2& 0.302 & 0.319 & 0.711 & 0.864 & 0.972 & -0.0059\\
(\ref{PL-wf}) n=3& 0.430 & 0.305 & 0.710 & 0.866 & 0.973 & -0.0061\\
(\ref{Tez91-wf}) & 0.028 & 0.301 & 0.711 & 0.865 & 0.973 & -0.0061\\
\end{tabular}
\end{table}
The calculation of the non-relativistic magnetic moments using
Eq.~(\ref{stat}) gives the following result which does not
depend on the model wave functions:  
\begin{equation}
\mu_{\rho\,NR} = 1 + \kappa_u + \kappa_{\bar d}\;.  
\label{muNR}
\end{equation} 
The values of parameters being fixed before we
obtain $\mu_{\rho\,NR}$ = 1.09.

The results of the calculation of the
$\rho$ -- meson statical moments are given in the Table I.
The relativistic results for the same parameters but without Wigner spin
rotation ($\omega_1 = \omega_2$ = 0 in (A1)--(A3)),
as well as of non-relativistic calculation are given, too.
The number of significant digits is
chosen so as to demonstrate the extent of the model
dependence of the calculations.

As one can see from the table the contributions of the spin
rotation to the magnetic moment and to MSR depend weakly on the
model for the quarks interaction in
$\rho$ meson. This contribution to the MSR is
24\%--26\% and is negative. This result differs from that of the
paper \cite{Kei94} where the contribution of spin rotation
(Melosh rotation) to MSR calculated in the frame of
light--front dynamics is positive.

\begin{figure}[htbp]
\epsfxsize=0.9\textwidth
\centerline{\psfig{figure=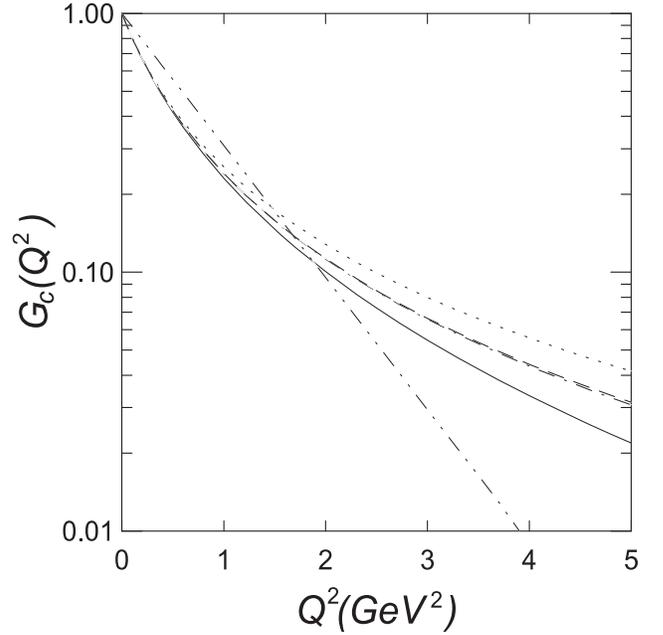,height=8.6cm,width=8.6cm}}
\vspace{0.3cm}
\caption{The results of the calculations of the
$\rho$-meson charge form factor with different model wave
functions. The values of the parameters are given in the
Table I. The solid line represents the relativistic calculation with the wave
function (\ref{HO-wf}), the dashed line -- with (\ref{PL-wf})
for $n =$ 3, dash--dot--line -- with (\ref{Tez91-wf}), dotted
line -- with (\ref{PL-wf}) for $n =$ 2, dot--dot--dash-line --
the non-relativistic calculation with (\ref{HO-wf})}
\label{fig:1} 
\end{figure}

\begin{figure}[htbp]
\epsfxsize=0.9\textwidth
\centerline{\psfig{figure=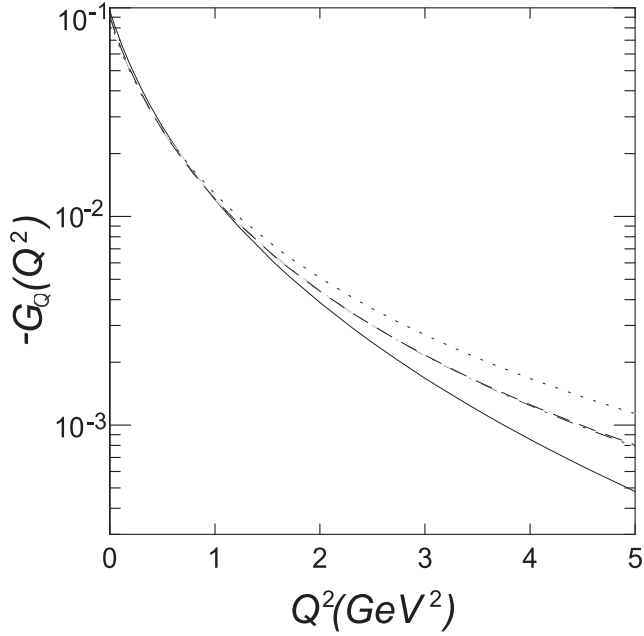,height=8.6cm,width=8.6cm}}
\vspace{0.3cm}
\caption{The results of the calculations of the
$\rho$-meson quadrupole form factor with different model wave
functions, legend as in Fig.1}
\label{fig:2}
\end{figure}

\begin{figure}[htbp]
\epsfxsize=0.9\textwidth
\centerline{\psfig{figure=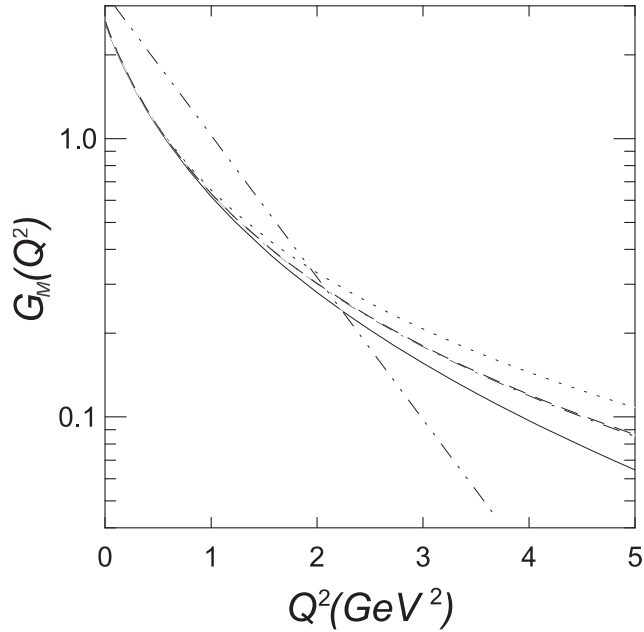,height=8.6cm,width=8.6cm}}
\vspace{0.3cm}
\caption{The results of the calculations of the
$\rho$-meson magnetic form factor with different model wave
functions, legend as in Fig.1}
\label{fig:3}
\end{figure}

\begin{figure}[htbp]
\epsfxsize=0.9\textwidth
\centerline{\psfig{figure=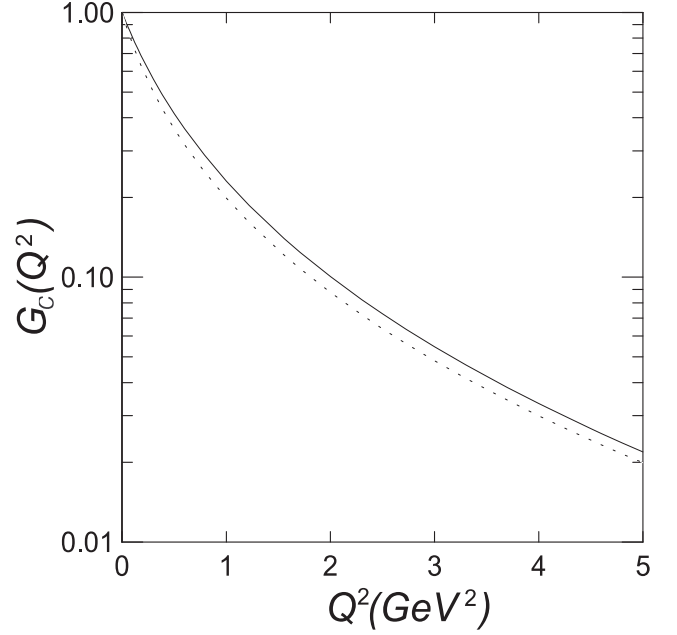,height=8.6cm,width=8.6cm}}
\vspace{0.3cm}
\caption{The contribution of the relativistic spin rotation
effect. The results of the relativistic calculation of the
$\rho$-meson charge form factor with the wave function
(\ref{HO-wf})  using the same parameters as in Fig.1.
Solid line represents the relativistic calculation with spin
rotation, dotted line -- the relativistic calculation without spin
rotation.}
\label{fig:4}
\end{figure}
The spin rotation
contribution to the magnetic moment in our calculations is
11\%--12\% and is negative, too.
The total relativistic
corrections to MSR in our approach are positive and enlarge the
non-relativistic value essentially -- almost twice in the
case of the model
(\ref{HO-wf}) and for 70\% -- 80\% for the models (\ref{PL-wf}),
(\ref{Tez91-wf}).

The total relativistic corrections for the
magnetic moment as compared to the non-relativistic result
(see Eq.~(\ref{muNR})) are negative and have the value of
21\%--22\%.  Let us note, that in the light--front dynamics
approach \cite{CaG95pl} a different result was obtained: the
positive relativistic correction of the value of 10\% to the
magnetic moment.

Let us note especially that in our calculation of the
$\rho$ -- meson quadrupole moment no ambiguities connected
with rotation symmetry arises -- in contrast with the
light--front dynamic calculations \cite{Kei94,CaG95pl}.

The results of calculations for the $\rho$ -- meson electromagnetic form
factors are represented in Fig.1--3.

Let us note that our charge form factor has no dip in contrast with
the results of the papers \cite{Kei94,CaG95pl}.
This principal difference is probably due to the effects of
rotation symmetry breaking in Refs.~\cite{Kei94,CaG95pl}.
The relativistic corrections in our approach diminish
essentially the rate of the decreasing of the charge and
magnetic form factors at large values of momentum transfer. We
demonstrate in the figures the case of the model (\ref{HO-wf})
with the exponential decreasing of the nonrelativistic form
factors with the increasing $Q^2$. The nonrelativistic
quadrupole form factor is zero in the absence of the $D$--state
in the two--particle system.

In Fig.4 the contribution of Wigner rotation of quark spins
to the $\rho$ -- meson charge form factor is shown. This
contribution depends weakly on the momentum transfer in the
range from 1 to  5 GeV$^2$ and its value is approximately 10\%.
In our calculation the sign of this contribution
differs from that obtained in the light--front dynamics approach
\cite{Kei94}. Let us note that similar difference takes place in
the case of pion electromagnetic structure, too (see
calculations in Ref.~\cite{ChC88pl} and Ref.~\cite{KrT99}).

\section{Conclusion}

The method of construction of the electromagnetic current matrix
elements for the relativistic two--particle composite systems
with nonzero total angular momentum is developed in the frame of
the instant form of RHD.

The method makes use of the Wigner--Eckart theorem on the
Poincar\'e group. It enables one to extract from the matrix
elements the reduced matrix elements -- invariant form
factors -- which in the case of composite systems are
generalized functions.

The obtained current operator matrix elements satisfy the
Lorentz--covariance condition and the conservation law.

The modified impulse approximation --- with the physical content
of the relativistic impulse approximation --- is formulated in
terms of reduced matrix elements. MIA conserves
Lorentz covariance of electromagnetic current and the
current conservation law.

The developed formalism is used to obtain a reasonable
description of the static moments and the electromagnetic form
factors of $\rho$ meson. A number of relativistic effects are
obtained, for example, the nonzero quadrupole moment (in the
case of $S$ state) due to the relativistic Wigner spin rotation.

So, in conclusion, it is shown that the instant form
of RHD can be used to obtain an adequate description of
the electroweak properties of composite systems with nonzero
total angular momentum.

\section*{Acknowledgements}

This work was supported in part by the Program "Russian
Universities--Basic Researches" (Grant No. 02.01.013).


\begin{flushleft}
{\bf Appendix 1}
\end{flushleft}

The charge form factor for free two--particle system:
$$
g_{0C}(s, Q^2, s') = \frac{1}{3}\,R(s, Q^2, s')\,Q^2\,
$$
$$
\times\left\{(s + s'+ Q^2)\left[G^u_E(Q^2)+G^{\bar d} _E(Q^2)\right]\,\right.
$$
$$
\times\left[2\,\cos(\omega_1-\omega_2) + \cos(\omega_1+\omega_2)\right]
$$
$$
-\,\frac{1}{M}\xi(s,Q^2,s')\left[G^u_M(Q^2)+G^{\bar d}_M(Q^2)\right]\,
$$
$$
\left.\times
\left[2\,\sin(\omega_1-\omega_2) - \sin(\omega_1+\omega_2)\right]\right\}\;,
\eqno{(A1)}
$$
The quadrupole form factor for free two--particle system:
$$
g_{0Q}(s, Q^2, s') = \frac{1}{2}\,R(s, Q^2, s')\,Q^2\,
$$
$$
\times\left\{(s + s'+ Q^2)\left[G^u_E(Q^2)+G^{\bar d} _E(Q^2)\right]\,\right.
$$
$$
\times\left[\cos(\omega_1-\omega_2) - \cos(\omega_1+\omega_2)\right]
$$
$$
- \,\frac{1}{M}\xi(s,Q^2,s')\left[G^u_M(Q^2)+G^{\bar d}_M(Q^2)\right]\,
$$
$$
\left.\times
\left[\sin(\omega_1-\omega_2) + \sin(\omega_1+\omega_2)\right]\right\}\;,
\eqno{(A2)}
$$
The magnetic form factor for free two--particle system:
$$
g_{0M}(s, Q^2, s') = -\,{2}\,R(s, Q^2, s')\,
$$
$$
\times\left\{\xi(s,Q^2,s')\left[G^u_E(Q^2)+G^{\bar d} _E(Q^2)\right]\,
\sin(\omega_1-\omega_2)
\right.
$$
$$
+ \frac{1}{4\,M}\left[G^u_M(Q^2)+G^{\bar d}_M(Q^2)\right]\,
\left[(s + s' +Q^2)\,Q^2\,\right.
$$
$$
\times\left(\frac{3}{2}\,\cos(\omega_1-\omega_2) +
\frac{1}{2}\cos(\omega_1+\omega_2)\right)
$$
$$
- \frac{1}{4}\xi(s,Q^2,s')
$$
$$
\times\left[\frac{(\sqrt{s'}+2\,M)(s-s'+Q^2)+(s'-s+Q^2)\sqrt{s'}}
{\sqrt{s'}(\sqrt{s'}+2\,M)}  \right.
$$
$$
\left. + \frac{(\sqrt{s}+2\,M)(s'-s+Q^2)+(s-s'+Q^2)\sqrt{s}}
{\sqrt{s}(\sqrt{s}+2\,M)}\right]
$$
$$
\times\left[\sin(\omega_1-\omega_2) - \sin(\omega_1+\omega_2)\right]
$$
$$
- \frac{1}{2}\xi^2(s,Q^2,s')\left[
\frac{1}{\sqrt{s'}(\sqrt{s'}+2\,M)}
+ \frac{1}{\sqrt{s}(\sqrt{s}+2\,M)}\right]
$$
$$
\left.\left.
\times
\left[\cos(\omega_1-\omega_2) + \cos(\omega_1+\omega_2)\right]\right]\right\}
\;.
\eqno{(A3)}
$$
Here
$$
R(s, Q^2, s') = \frac{(s + s'+ Q^2)}{2\sqrt{(s-4M^2) (s'-4M^2)}}\,
$$
$$
\times\frac{\vartheta(s,Q^2,s')}{{[\lambda(s,-Q^2,s')]}^{3/2}}
\frac{1}{\sqrt{1+Q^2/4M^2}}\;,
$$
$$
\xi(s,Q^2,s')=\sqrt{ss'Q^2-M^2\lambda(s,-Q^2,s')}\;,
$$
$\omega_1$ ¨ $\omega_2$ -- the Wigner rotation parameters:
$$
\omega_1 =
\arctan\frac{\xi(s,Q^2,s')}{M\left[(\sqrt{s}+\sqrt{s'})^2 +
Q^2\right] + \sqrt{ss'}(\sqrt{s} +\sqrt{s'})}\;,
$$
$$
\omega_2 = \arctan\frac{
\alpha (s,s') \xi(s,Q^2,s')} {M(s + s' + Q^2)
\alpha (s,s')
+ \sqrt{ss'}(4M^2 + Q^2)}\;,
$$
here $\alpha (s,s') = 2M + \sqrt{s} + \sqrt{s'} $,
$\vartheta(s,Q^2,s')=
\theta(s'-s_1)-\theta(s'-s_2)$, $\theta$ - the step--function.
$$
s_{1,2}=2M^2+\frac{1}{2M^2} (2M^2+Q^2)(s-2M^2)
$$
$$
\mp \frac{1}{2M^2}
\sqrt{Q^2(Q^2+4M^2)s(s-4M^2)}\;.
$$
$M$ -- the mass of $u$-- and $\bar d$ quarks. The functions
$s_{1,2}(s,Q^2)$ give the kinematically available region
in the plane $(s,s')$.  $G^{u,\bar d}_{E,M}(Q^2)$-- Sachs form factors of
$u$-- and $\bar d$ quarks.

\end{document}